\documentclass[final,5p,times,twocolumn]{elsarticle}

\usepackage{amssymb}
\usepackage[utf8]{inputenc}
\usepackage{fixltx2e}
\usepackage{caption}
\usepackage{subcaption}
\usepackage[multiple]{footmisc}
\usepackage{graphicx}
\usepackage{amsmath}
\usepackage{tabularx}
\usepackage{multirow}
\usepackage{booktabs}
\usepackage{listings}
\usepackage{natbib}
\usepackage{tikz}
\usepackage{bm}
\usepackage{array}
\usepackage[htt]{hyphenat}
\usepackage{xcolor}
\usepackage{float}
\usepackage{tcolorbox}
\usepackage{url}

\usepackage[edges]{forest}
\usetikzlibrary{shadows}

\usepackage{flexisym}

\usepackage{pifont}

\newcommand{\cmark}{\ding{51}}%
\newcommand{\xmark}{\ding{55}}%

\newcolumntype{P}[1]{>{\centering\arraybackslash}p{#1}}
\newcolumntype{M}[1]{>{\centering\arraybackslash}m{#1}}
\newcolumntype{Y}{>{\centering\arraybackslash}X}

\usepackage{makecell}
\usepackage[raggedrightboxes]{ragged2e}
\usepackage{listings}
    \lstdefinelanguage{diff}{
    basicstyle=\rmfamily\small,
    morecomment=[f][\color{diffstart}]{@@},
    morecomment=[f][\color{diffincl}]{+\ },
    morecomment=[f][\color{diffrem}]{-\ },
    }

\newcommand{\tufanoData}{Tufano\textsubscript{data}}
\newcommand{\CodeReviewerData}{CodeReviewer\textsubscript{data}}

\newcommand{\DACTData}{D-ACT\textsubscript{data}}

\newcommand{\ChatGPTZeroShot}{GPT-3.5\textsubscript{Zero-shot}}

\newcommand{\ChatGPTFewShot}{GPT-3.5\textsubscript{Few-shot}}

\newcommand{\ChatGPTFineTune}{GPT-3.5\textsubscript{Fine-tuned}}

\newcommand{\ea}{{\em et al.}}

\newcommand{\smallsection}[1]{\textbf{#1. }}
\newcommand{\revised}[2]{{\color{black}{#2}}}

\newcommand{\revisedTwo}[2]{{\color{black}{#2}}}

\newcommand{\rqone}{What is the most effective approach to leverage LLMs for code review automation?}

\newcommand{\rqtwo}{What is the benefit of model fine-tuning on GPT-3.5 for code review automation?}

\newcommand{\rqthree}{What is the most effective prompting strategy on GPT-3.5 for code review automation?}

\begin{document}

\title{Fine-Tuning and Prompt Engineering for Large Language Models-based Code Review Automation}

\author[inst1]{Chanathip Pornprasit}
\ead{chanathip.pornprasit@monash.edu}

\author[inst1]{Chakkrit Tantithamthavorn\corref{cor1}}
\ead{chakkrit@monash.edu}

\affiliation[inst1]{organization={Monash University},
            country={Australia}}

\cortext[cor1]{Corresponding author.}

\begin{abstract}

\textbf{Context}: The rapid evolution of Large Language Models (LLMs) has sparked significant interest in leveraging their capabilities for automating code review processes. 
Prior studies often focus on developing LLMs for code review automation, yet require expensive resources, which is infeasible for organizations with limited budgets and resources.
Thus, fine-tuning and prompt engineering are the two common approaches to leveraging LLMs for code review automation.

\indent\textbf{Objective}: We aim to investigate the performance of LLMs-based code review automation based on two contexts, i.e., when LLMs are leveraged by fine-tuning and prompting.
Fine-tuning involves training the model on a specific code review dataset, while prompting involves providing explicit instructions to guide the model's generation process without requiring a specific code review dataset. 


\indent\textbf{Method}: 
We leverage model fine-tuning and inference techniques (i.e., zero-shot learning, few-shot learning and persona) on LLMs-based code review automation.
In total, we investigate 12 variations of two LLMs-based code review automation (i.e., GPT-3.5 and Magicoder), and compare them with the Guo~\ea's approach and three existing code review automation approaches (i.e., CodeReviewer, TufanoT5 and D-ACT).

\indent\textbf{Results}:
The fine-tuning of GPT 3.5 with zero-shot learning helps GPT-3.5 to achieve 73.17\% -74.23\% higher EM than the Guo~\ea's approach.
In addition, when GPT-3.5 is not fine-tuned, GPT-3.5 with few-shot learning achieves 46.38\% - 659.09\% higher EM than GPT-3.5 with zero-shot learning.

\indent\textbf{Conclusions}: Based on our results, we recommend that (1) LLMs for code review automation should be fine-tuned to achieve the highest performance.;
and (2) when data is not sufficient for model fine-tuning (e.g., a cold-start problem), few-shot learning without a persona should be used for LLMs for code review automation.
Our findings contribute valuable insights into the practical recommendations and trade-offs associated with deploying LLMs for code review automation.

\end{abstract}


\maketitle


\section{Introduction}





Code review is a software quality assurance practice where developers other than an author (aka. reviewers) review a code change that the author creates to ensure the quality of the code change before being integrated into a codebase.
While code review can ensure high software quality, code review is still time-consuming and expensive.
Thus, neural machine translation (NMT)-based code review automation approaches were proposed~\cite{tufano2019learning, tufano2021towards, AutoTransform} to facilitate and expedite the code review process.
However, prior studies~\cite{pornprasit2023d, li2022automating} found that 
such approaches are still not perfect due to limited knowledge of the NMT-based code review automation models that are trained on a small code review dataset.

To address the aforementioned challenge of the NMT-based code review automation approaches, recent work proposed large language model (LLM)-based approaches for the code review automation task~\cite{li2022automating, tufano2022using}.
A large language model is a large deep learning model that is based on the transformer architecture~\cite{Vaswani2017} and pre-trained on massive textual datasets.
An example of the LLM-based code review automation approaches includes CodeReviewer~\cite{li2022automating}, a pre-trained LLM that is based on the CodeT5~\cite{wang2021codet5} model.
Li~\ea~\cite{li2022automating} showed that their proposed CodeReviewer outperforms prior NMT-based code review automation approaches~\cite{tufano2019learning, tufano2021towards, AutoTransform}.
However, the training process of CodeReviewer requires a lot of computing resources (i.e., two DGX-2 servers equipped with 32 NVIDIA V100 GPUs in total).
Such large computing resources are infeasible for organizations with limited budgets.




Since pre-training LLMs for code review automation can be expensive, fine-tuning and prompt engineering are the two common approaches to leveraging LLMs for code review automation.
In particular, fine-tuning involves further training LLMs that are already pre-trained on a specific code review dataset.
For example, Lu~\ea~\cite{llamareviewer} proposed LLaMa-Reviewer, which is the LLM-based code review automation approach that is being fine-tuned on a base LLaMa model~\cite{touvron2023llama}.
On the other hand, prompting~\cite{gao2023makes, gao2023constructing, white2023prompt} involves providing explicit instructions to guide the model’s generation process without requiring a specific code review dataset.
For instance, Guo~\ea~\cite{guo2023exploring} conducted an empirical study to investigate the potential of GPT-3.5 for code review automation by using zero-shot learning with GPT-3.5.

While Guo~\ea~\cite{guo2023exploring} demonstrate the potential of using GPT-3.5 for code review automation, their study still has the following limitations.
First, the results of Guo~\ea~\cite{guo2023exploring} are limited to zero-shot GPT-3.5.
However, there are other approaches to leverage GPT-3.5 (i.e., fine-tuning and few-shot learning) that are not included in their study.
Thus, it is difficult for practitioners to conclude which approach is the best for leveraging LLMs for code review automation.
Second, although prior studies~\cite{chowdhery2023palm, wei2021finetuned, chen2021evaluating} found that model fine-tuning can improve the performance of pre-trained LLMs, Guo~\ea~\cite{guo2023exploring} did not evaluate the performance of LLMs when being fine-tuned.
Thus, it is difficult for practitioners to conclude whether LLMs for code review automation should be fine-tuned to achieve the most effective results.
Third, Guo~\ea~\cite{guo2023exploring} did not investigate the impact of few-shot learning, which can improve the performance of LLMs over zero-shot learning~\cite{brown2020language, kang2023large, geng2024large}.
Hence, it is difficult for practitioners to conclude which prompting strategy (i.e., zero-shot learning, few-shot learning, and a persona) is the most effective for code review automation.


\revised{R1-1}{
In this work, we aim to investigate the performance of LLMs-based code review automation based on two contexts, i.e., when LLMs are leveraged by fine-tuning and prompting.
}
In particular, we evaluate two LLMs (i.e., GPT-3.5 and Magicoder~\cite{wei2023magicoder}) and the existing LLM-based code review automation approaches~\cite{pornprasit2023d, li2022automating, tufano2022using} with respect to the following evaluation measures: Exact Match (EM)~\cite{tufano2019learning, pornprasit2023d} and CodeBLEU~\cite{ren2020codebleu}.
Through the experimental study of the three code review automation datasets (i.e., Code\-Reviewer\textsubscript{data}~\cite{li2022automating}, \tufanoData~\cite{tufano2022using}, and \DACTData~\cite{pornprasit2023d}), we answer the following three research questions:

\textbf{(RQ1) \rqone}

\smallsection{Result}
The fine-tuning of GPT 3.5 with zero-shot learning achieves 73.17\% -74.23\% higher EM than the Guo~\ea's approach~\cite{guo2023exploring} (i.e., GPT 3-5 without fine-tuning).
The results imply that GPT-3.5 should be fine-tuned to achieve the highest performance.


\textbf{(RQ2) \rqtwo}

\smallsection{Result}
The fine-tuning of GPT 3.5 with few-shot learning achieves 63.91\% - 1,100\% higher Exact Match than those that are not fine-tuned.
The results indicate that fine-tuned GPT-3.5 can generate more correct revised code than GPT-3.5 without fine-tuning.




\textbf{(RQ3) \rqthree}

\smallsection{Result}
GPT-3.5 with few-shot learning achieves 46.38\% - 659.09\% higher Exact Match than GPT-3.5 with zero-shot learning.
On the other hand, when a persona is included in input prompts, GPT-3.5 achieves 1.02\% - 54.17\% lower Exact Match than when the persona is not included in input prompts.
The results indicate that the best prompting strategy when using GPT-3.5 without fine-tuning is using few-shot learning without a persona.





\textbf{Recommendation.}
Based on our results, we recommend that 
(1) LLMs for code review automation should be fine-tuned to achieve the highest performance;
and (2) when data is not sufficient for model fine-tuning (e.g., a cold-start problem), few-shot learning without a persona should be used for LLMs for code review automation.


\textbf{Contributions.}
In summary, the main contributions of our work are as follows:

\begin{itemize}
    \item We are the first to investigate the performance of LLMs-based code review automation when using model fine-tuning and inference techniques (i.e., zero-shot learning, few-shot learning and persona).
    \item We provide recommendations for adopting 
    LLMs for code review automation to practitioners.
\end{itemize}


\textbf{Open Science.}
Our fine-tuned models, script and results are made available online~\cite{github-repo}.

\textbf{Paper Organization.}
Section~\ref{sec:background} describes the related work and formulates research questions.
Section~\ref{sec:study-design} describes the study design of our study.
Section~\ref{sec:results} presents the experiment results.
Section~\ref{sec:discussion} discusses our experiment results.
Section~\ref{sec:threat} describes possible threats to the validity.
Section~\ref{sec:conclusion} draws the conclusions of our work.

\section{Related Work and Research Questions}  \label{sec:background}


\begin{figure}
    \includegraphics[width=\columnwidth]{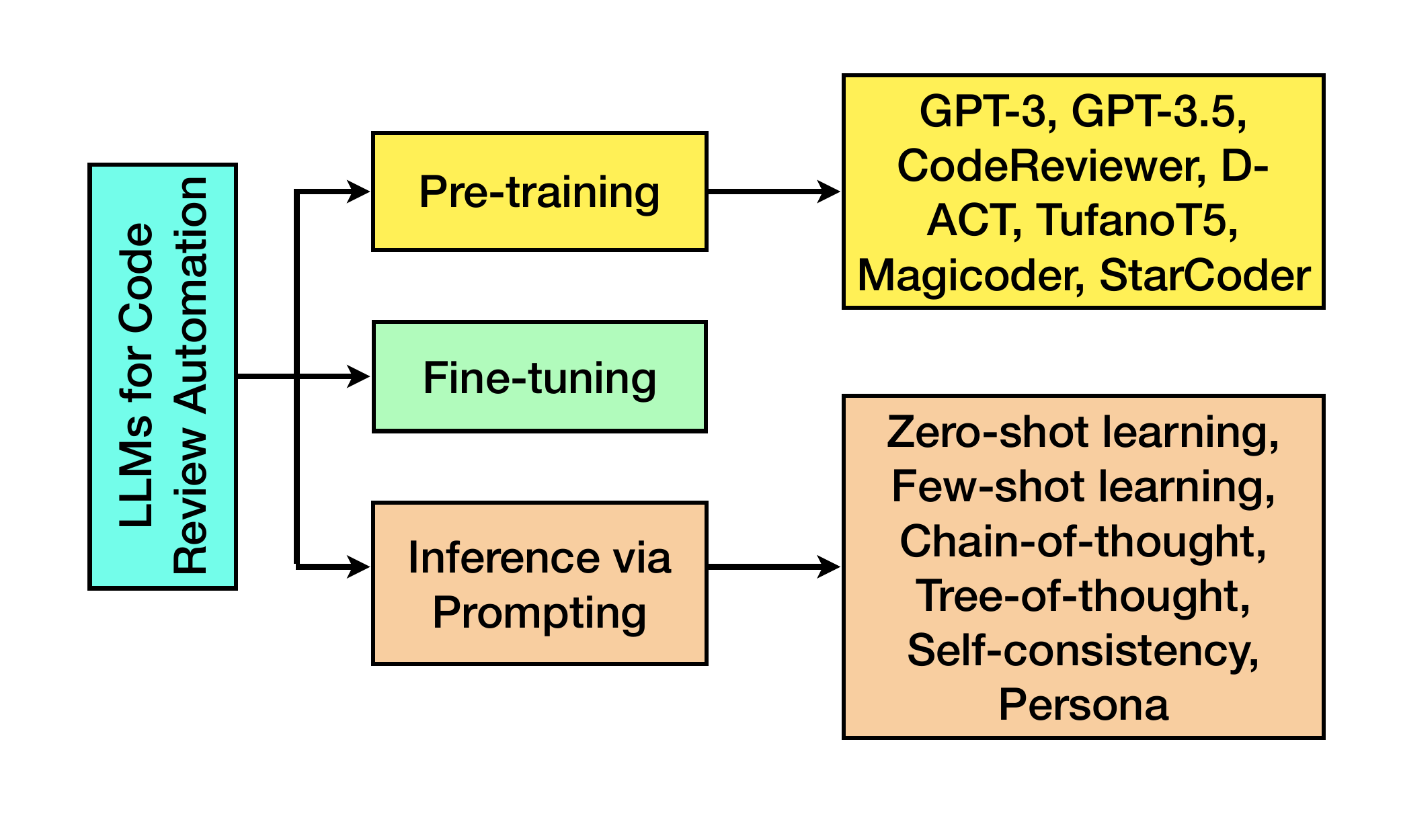}
    \caption{An overview of the modelling pipeline of LLMs for code review automation.}
    \label{fig:LLM-modeling-pipeline}
\end{figure}

In this section, we provide the background knowledge of code review automation, discuss the existing large-language model-based code review automation approaches, and formulate the research questions.

\subsection{Code Review Automation}



Code review is a software quality assurance practice where developers other than an author (aka. reviewers) provide feedback for a code change created by the author to ensure that the code change has sufficient quality to meet quality standards.
While code review can ensure high software quality, code review is still time-consuming and expensive.
Thus, developers still face challenges in receiving timely feedback from reviewers~\cite{macleod2017code, rigby2013convergent}.
Therefore, code review automation approaches~\cite{ tufano2019learning, pornprasit2023d, li2022automating, tufano2022using, tufan2021towards} were proposed to facilitate and expedite the code review process.


Code review automation is generally formulated as a sequence generation task, where a language model is trained to learn the relationship between the submitted code and the revised code. 
Then, during the inference phase, the model aims to generate a revised version of a code change.
Recently, neural machine translation (NMT)-based code review automation approaches were proposed~\cite{tufano2019learning, tufano2021towards, AutoTransform}.
Typically, NMT-based code review automation approaches are trained on a specific code review dataset. 
However, prior studies~\cite{pornprasit2023d, li2022automating} found that NMT-based code review automation approaches can perform well, but is still not perfect.
This imperfect performance has to do with the limited knowledge of the NMT-based code review automation approaches that are being trained on a small code review dataset.

\subsection{LLMs-based Code Review Automation Approaches}

Large language models (LLMs) for code review automation refer to large language models specifically designed to support code review automation tasks, aiming to understand and generate source code written by developers and natural languages written by reviewers. 
Since source code and comments often have their own semantic and syntactical structures, recent work proposed various LLMs-based code review automation approaches~\cite{li2022automating, tufano2022using, 
 llamareviewer}. 
For example, Li~\ea~\cite{li2022automating} proposed CodeReviewer, a pre-trained LLM that is based on the CodeT5 model~\cite{wang2021codet5}.
Prior studies found that LLMs-based code review automation approaches often outperform NMT-based ones~\cite{pornprasit2023d, li2022automating, tufano2022using}.
For example, Li~\ea~\cite{li2022automating} found that their proposed approach outperforms a transformer-based NMT model by 11.76\%.
Below, we briefly discuss the general modelling pipeline of LLMs for code review automation presented in Figure~\ref{fig:LLM-modeling-pipeline}.

\textbf{Model Pre-Training} refers to the initial phase of training a large language model, where the model is exposed to a large amount of unlabeled data to learn general language representations.
This phase aims to initialize the model's parameters and learn generic features that can be further fine-tuned for specific downstream tasks.
Recently, there have been many large language models for code (i.e., LLMs that are specifically trained on source code and related natural languages).
For example, the open-source community-developed large language models such as Code-LLaMa~\cite{roziere2023codeLLaMa}, StarCoder~\cite{li2023starcoder}, and Magicoder~\cite{wei2023magicoder}; and the commercial large language models such as GPT-3.5.

However, the development of large language models for code requires expensive GPU resources and budget. 
For example, GPT-3.5 requires 10,000 NVIDIA V-100 GPUs for model pre-training.\footnote{https://gaming.lenovo.com/emea/threads/17314-The-hardware-behind-ChatGPT} 
LLaMa2~\cite{touvron2023llama2} requires Meta’s Research Super Cluster (RSC) as well as internal production clusters, which consists of approximately 2,000 NVIDIA A-100 GPUs in total.
Therefore, many software organizations with limited resources and budgets may not be able to develop their large language models.
Thus, fine-tuning and prompt engineering are the two common approaches to leverage the existing LLMs for code review automation when expensive GPU resources are not available for pre-training a large language model from scratch, where these techniques are more desirable for many organizations to quickly adopt new technologies.


\textbf{Model Fine-Tuning} is a common practice, particularly in transfer learning scenarios, where a model pre-trained on a large dataset (source domain, e.g., source code understanding) is adapted to a related but different task or dataset (target domain, e.g., code review automation).
Recently, researchers have leveraged model fine-tuning techniques for LLMs to improve the performance of code review automation approaches.
For example, Lu~\ea~\cite{llamareviewer} proposed LLaMa-Reviewer, which is an LLM-based code review automation approach that is being fine-tuned on a base LLaMa model~\cite{touvron2023llama} using three code review automation tasks, i.e., a review necessity prediction task to check if diff hunks need a review, a code review comment generation task to generates pertinent comments for a given code snippet, and a code refinement task to generate minor adjustment to the existing code.
Lu~\ea~\cite{llamareviewer} found that the fine-tuning step on LLMs can greatly improve the performance of the existing code review automation approaches.


\textbf{Inference} refers to the process of using a pre-trained language model to generate source code based on a given natural language prompt instruction.
Therefore, prompt engineering plays an important role in leveraging LLMs for code review automation to guide LLMs to generate the desired output. 
Different prompting strategies have been proposed.\footnote{https://www.promptingguide.ai/}
For example, zero-shot learning, few-shot learning~\cite{brown2020language, liu2021makes, wei2023larger}, chain-of-thought~\cite{kim2023cot, wei2022chain}, tree-of-thought~\cite{kim2023cot, wei2022chain}, self-consistency~\cite{wang2022self}, and persona~\cite{white2023prompt}.
Nevertheless, not all prompting strategies are relevant to code review automation.
For example, chain-of-thought, self-consistency and tree-of-thought promptings are not applicable to the code review automation task since they are designed for arithmetic and logical reasoning problems.
Thus, we exclude them from our study.

In contrast, zero-shot learning, few-shot learning, and persona prompting are the instruction-based prompting strategies, which are more suitable for software engineering (including code review automation) tasks~\cite{xu2023prompting, dilhara2024unprecedented, misu2024towards, jiang2023llmparser}.
In particular, zero-shot learning involves prompting LLMs to generate an output from a given instruction and an input.
On the other hand, few-shot learning~\cite{brown2020language, liu2021makes, wei2023larger} involves prompting LLMs to generate an output from $N$ demonstration examples $\{(x_1, y_1), (x_2, y_2), ..., (x_N, y_N)\}$ and an actual input in a testing set, where $x_i$ and $y_i$ are the inputs and outputs obtained from a training set, respectively.
Persona~\cite{white2023prompt} involves prompting LLMs to act as a specific role or persona to ensure that LLMs will generate output that is similar to the output generated by a specified persona.

\begin{table}[htbp]
  \centering
  \caption{The differences between our work and Guo~\ea's work~\cite{guo2023exploring}.}
  \resizebox{\columnwidth}{!}{
    \begin{tabular}{|p{0.15\textwidth}|p{0.2\textwidth}|p{0.2\textwidth}|}
\cline{2-3} \multicolumn{1}{r|}{}  & Guo~\ea~\cite{guo2023exploring} & Our work \\
    \hline
    LLMs/approaches & GPT-3.5, CodeReviewer~\cite{li2022automating} & GPT-3.5, Magicoder~\cite{wei2023magicoder}, CodeReviewer~\cite{li2022automating}, TufanoT5~\cite{tufano2022using}, D-ACT~\cite{pornprasit2023d} \\
    \hline
    Include fine-tuning LLMs? & No    & \multicolumn{1}{l|}{Yes} \\
    \hline
    Prompting techniques & Zero-shot learning, Persona & Zero-shot learning, Few-shot learning, Persona \\
    \hline
    
    \end{tabular}%
    }
  \label{tab:work-comparison}%
\end{table}%

\subsection{GPT-3.5 for Code Review Automation}

Recently, Guo~\ea~\cite{guo2023exploring} conducted an empirical study to investigate the potential of GPT-3.5 for code review automation.
However, their study still has the following limitations.


\textbf{First, the results of Guo~\ea~\cite{guo2023exploring} are limited to zero-shot GPT-3.5.}
In particular, Guo~\ea~\cite{guo2023exploring} conducted experiments to find the best prompt for leveraging zero-shot learning with GPT-3.5.
However, there are other approaches to leverage GPT-3.5 (i.e., fine-tuning and few-shot learning) that are not included in their study.
The lack of a systematic evaluation of the use of fine-tuning and few-shot learning on GPT-3.5 makes it difficult for practitioners to conclude which approach is the best for leveraging LLMs for code review automation.
To address this challenge, we formulate the following research question.


\begin{tcolorbox}[left=2pt, right=2pt, top=2pt,bottom=2pt]
RQ1: \rqone
\end{tcolorbox}

\textbf{Second, the performance of LLMs when being fine-tuned is still unknown.}
In particular, Guo~\ea~\cite{guo2023exploring} did not evaluate the performance of LLMs when being fine-tuned.
However, prior studies~\cite{chowdhery2023palm, wei2021finetuned, chen2021evaluating} found that model fine-tuning can improve the performance of pre-trained LLMs.
The lack of experiments with model fine-tuning makes it difficult for practitioners to conclude whether LLMs for code review automation should be fine-tuned to achieve the most effective results.
To address this challenge, we formulate the following research question.



\begin{tcolorbox}[left=2pt, right=2pt, top=2pt,bottom=2pt]
RQ2: \rqtwo
\end{tcolorbox}

\textbf{Third, the performance of LLMs for code review automation when using few-shot learning is still unknown.}
In particular, Guo~\ea~\cite{guo2023exploring} did not investigate the impact of few-shot learning on LLMs for code review automation.
However, recent work~\cite{brown2020language, kang2023large, geng2024large} found that few-shot learning could improve the performance of LLMs over zero-shot learning.
The lack of experiments with few-shot learning on LLMs for code review automation makes it difficult for practitioners to conclude which prompting strategy (i.e., zero-shot learning, few-shot learning, and persona) is the most effective for code review automation.
To address this challenge, we formulate the following research question.


\begin{tcolorbox}[left=2pt, right=2pt, top=2pt,bottom=2pt]
RQ3: \rqthree
\end{tcolorbox}

\begin{figure*}
    \centering
    \includegraphics[width=\linewidth]{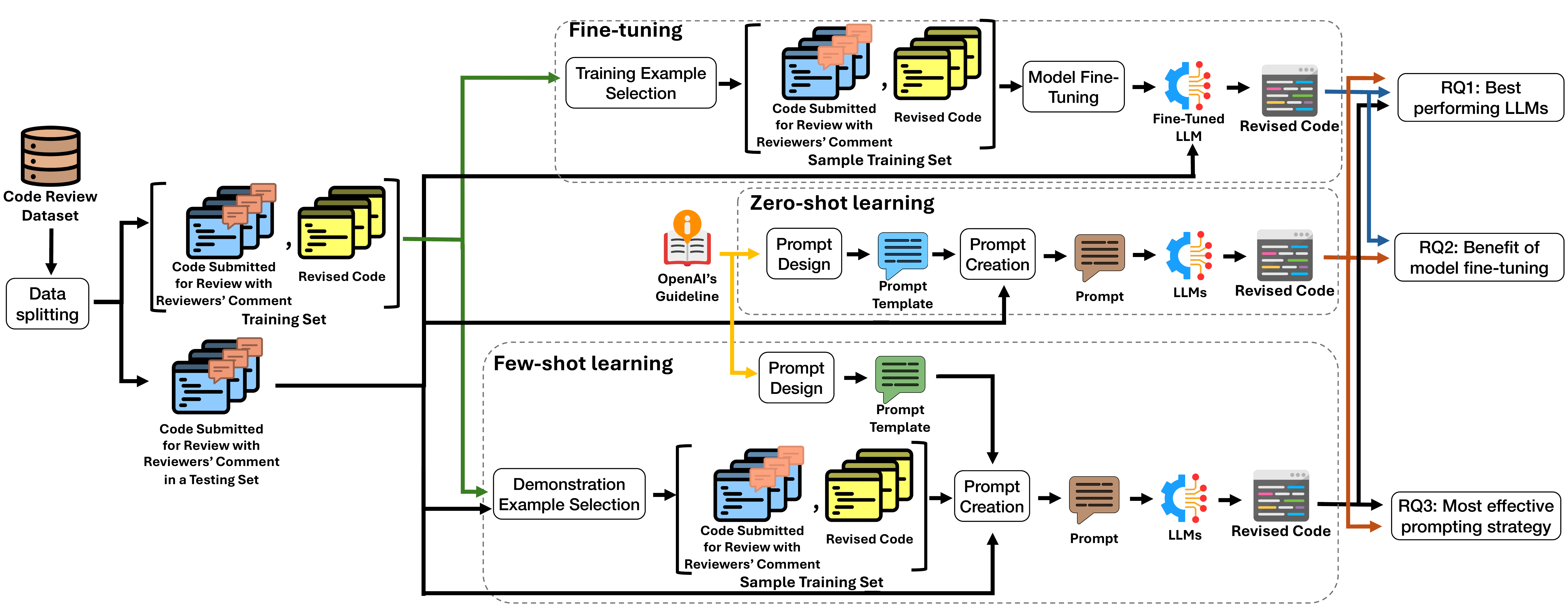}
    \caption{An overview of our experimental design (A persona is a part of zero-shot and few-shot learning).}
    \label{fig:study-design}
\end{figure*}

\begin{table}[htbp]
  \centering
  \caption{Experimental settings in our study. We do not include experimental settings \#3 and \#4 since LLMs already learn the relationship between input (i.e., code submitted for review) and output (i.e., revised code).}
    \begin{tabular}{|c|c|c|c|}
    \hline
    \multicolumn{1}{|c|}{\multirow{2}{*}{\makecell{Experimental\\setting}}} & \multirow{2}{*}{Fine-Tuning} & \multicolumn{2}{c|}{Inference Technique} \\
\cline{3-4}          &       & Prompting & \multicolumn{1}{l|}{Use Persona} \\
    \hline
    \#1   & \multirow{4}{*}{\cmark} & \multirow{2}{*}{Zero-shot} & \xmark \\
\cline{1-1}\cline{4-4}    \#2   &       &       & \cmark \\
\cline{1-1}\cline{3-4}    \#3   &       & \multirow{2}{*}{Few-shot} & \xmark \\
\cline{1-1}\cline{4-4}    \#4   &       &       & \cmark \\
    \hline
    \#5   & \multirow{4}{*}{\xmark} & \multirow{2}{*}{Zero-shot} & \xmark \\
\cline{1-1}\cline{4-4}    \#6   &       &       & \cmark \\
\cline{1-1}\cline{3-4}    \#7   &       & \multirow{2}{*}{Few-shot} & \xmark \\
\cline{1-1}\cline{4-4}    \#8   &       &       & \cmark \\
    \hline
    \end{tabular}%
  \label{tab:exp-setting}%
\end{table}%

\section{Experimental Design} \label{sec:study-design}




In this section, we provide an overview and details of our experimental design.

\subsection{Overview}
The goal of this work is to investigate which LLMs perform best when using model fine-tuning and inference techniques (i.e., zero-shot learning, few-shot learning~\cite{brown2020language, liu2021makes, wei2023larger}, and persona~\cite{white2023prompt}).
To achieve this goal, we conduct experiments with two LLMs (i.e., GPT-3.5 and Magicoder~\cite{wei2023magicoder}) on the following datasets that are widely studied in the code review automation literature~\cite{pornprasit2023d, llamareviewer, guo2023exploring, zhou2023generation}: \CodeReviewerData~\cite{li2022automating}, \tufanoData~\cite{tufano2022using} and \DACTData~\cite{pornprasit2023d}.
We use Magicoder~\cite{wei2023magicoder} in our experiment since it is further trained on high-quality synthetic instructions and solutions.

In this study, we conduct experiments under six settings as presented in Table~\ref{tab:exp-setting}.
According to the table, when the LLMs are fine-tuned, we use zero-shot learning with and without a persona.
We do not use few-shot learning with the fine-tuned LLMs since the LLMs already learn the relationship between an input (i.e., code submitted for review) and an output (i.e., improved code).
On the other hand, when the LLMs are not fine-tuned, we use zero-shot learning and few-shot learning, where each inference technique is used with and without a persona.
Finally, we conduct 36 experiments in total (2 LLMs $\times$ 6 settings $\times$ 3 datasets).


Figure~\ref{fig:study-design} provides an overview of our experimental design.
To begin, the studied code review datasets are split into training and testing sets.
The training set consists of the code submitted for review and reviewers' comments as input; and revised code as output.
On the other hand, the testing set consists of only code submitted for review and reviewers' comments.
Next, to fine-tune the studied LLMs, we first randomly obtain a set of training examples from the training set since using the whole training set is prohibitively expensive.
Then, we use the selected training examples to fine-tune the studied LLMs.
On the other hand, to use the inference techniques (i.e., zero-shot learning, few-shot learning and a persona), we first design prompt templates for each inference technique based on the guideline from OpenAI\footnote{https://help.openai.com/en/articles/6654000-best-practices-for-prompt-engineering-with-openai-api\label{fn:open-ai-guide-1}}\footnote{https://platform.openai.com/docs/guides/prompt-engineering/strategy-write-clear-instructions\label{fn:open-ai-guide-2}}.
However, since few-shot learning requires demonstration examples, we select a set of demonstration examples for each testing sample from the training set.
Then, we create prompts that look similar to the prompt templates.
Finally, we use the studied LLMs to generate revised code from given prompts.
We explain the details of the studied datasets, model fine-tuning, inference via prompting, evaluation measures, and hyper-parameter settings below.


\begin{table*}
  \centering
  \caption{A statistic of the studied datasets (the dataset of Android, Google and Ovirt are from the \DACTData~dataset~\cite{pornprasit2023d}).}
    \begin{tabular}{|c|c|c|c|c|c|c|}
\hline   Dataset  & \# Train & \# Validation & \# Test & \# Language & Granularity & Has Comment\\
    \hline
    \CodeReviewerData~\cite{li2022automating} & 150,405 & 13,102 & 13,104 & 9 & Diff Hunk & \cmark \\
    \tufanoData~\cite{tufano2022using} & 134,238 & 16,779 & 16,779 & 1 & Function & \cmark/\xmark \\
    Android~\cite{pornprasit2023d} & 14,690 & 1,836 & 1,835 & 1 & Function & \xmark \\
    Google~\cite{pornprasit2023d} & 9,899 & 1,237 & 1,235 & 1 & Function & \xmark \\
    Ovirt~\cite{pornprasit2023d}  & 21,509 & 2,686 & 2,688 & 1 & Function & \xmark \\
    \hline
    \end{tabular}%
  \label{tab:dataset}%
\end{table*}%

\subsection{The Studied Datasets}


Recently, Tufano~\ea~\cite{tufano2019learning, tufano2021towards} collected datasets with the constraint that revised code must not contain the code tokens (e.g., identifiers) that do not appear in code submitted for review.
Thus, such datasets do not align with the real code review practice since developers may add new code tokens when they revise their submitted code.
Therefore, in this study, we use the CodeReviewer~\cite{li2022automating}, TufanoT5~\cite{tufano2022using}, and D-ACT~\cite{pornprasit2023d} datasets, which do not have the above constraint in data collection instead.
The details of the studied datasets are as follows (the statistic of the studied datasets is presented in Table~\ref{tab:dataset}).

\begin{itemize}
        
        \item \textbf{\CodeReviewerData}: Li~\ea~\cite{li2022automating} collected this dataset from the GitHub projects across nine programming languages (i.e., C, C++, C\#, Java, Python, Ruby, php, Go, and Javascript).
        The dataset contains triplets of the code submitted for review (diff hunk granularity), a reviewer's comment, and the revised version of the code submitted for review (diff hunk granularity).
        
        \item \textbf{\tufanoData}: Tufano~\ea~\cite{tufano2022using} collected this dataset from Java projects in GitHub, and 6,388 Java projects hosted in Gerrit.
        Each record in the dataset contains a triplet of code submitted for review (function granularity), a reviewer's comment, and code after being revised (function granularity).
        Tufano~\ea~\cite{tufano2022using} created two types of this dataset (i.e., Tufano\textsubscript{data} (with comment) and \tufanoData~(without comment)).
        
        \item  \textbf{\DACTData}: Pornprasit~\ea~\cite{pornprasit2023d} collected this dataset from the three Java projects hosted on Gerrit (i.e., Android, Google and Ovirt).
        Each record in the dataset contains a triplet of codebase (function granularity), code of the first version of a patch (function granularity), and code of the approved version of a patch (function granularity).
    \end{itemize}

\subsection{Model Fine-Tuning}

To fine-tune the studied LLMs, as suggested by OpenAI, we first select a few training examples to fine-tune an LLM to see if the performance improves.
Thus, we randomly select a set of examples from the whole training set by using the \texttt{random} function in Python to reduce bias in the data selection.
However, there is no existing rule or principle to determine the number of examples that should be selected from a training set.
Thus, we use the trial-and-error approach to determine the suitable number of training examples.
To do so, we start by using approximately 6\% training examples from the whole training set to fine-tune GPT-3.5.
We find that GPT-3.5 that is fine-tuned with such training examples outperforms the existing code review automation approaches~\cite{pornprasit2023d, li2022automating, tufano2022using}.
Therefore, based on the above finding, we use 6\% training examples for the whole experiment.

After that, the selected training examples is used to fine-tune the studied LLMs.
In particular, we fine-tune GPT-3.5 by using the API provided by OpenAI\footnote{https://platform.openai.com/docs/guides/fine-tuning/create-a-fine-tuned-model}.
On the other hand, to fine-tune Magicoder~\cite{wei2023magicoder}, we leverage the state-of-the-art parameter-efficient fine-tuning technique called DoRA~\cite{liu2024dora}.

\begin{figure}[!h]
    \centering
    \begin{subfigure}{\columnwidth}
         \centering
         \includegraphics[width=\columnwidth, page = 1, trim = {0 3cm 0 0}, clip]{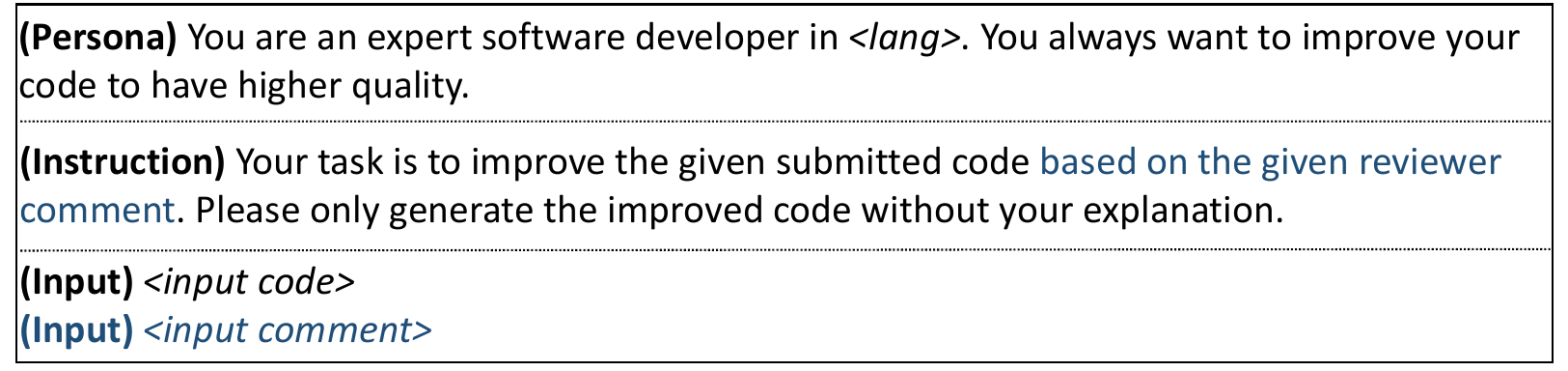}
         \caption{A prompt template for zero-shot learning.}
         \label{fig:zero-shot prompt}
     \end{subfigure}
     \par\bigskip
     \begin{subfigure}{\columnwidth}
         \centering
         \includegraphics[width=\columnwidth, page = 1, trim = {0 3cm 0 0}, clip]{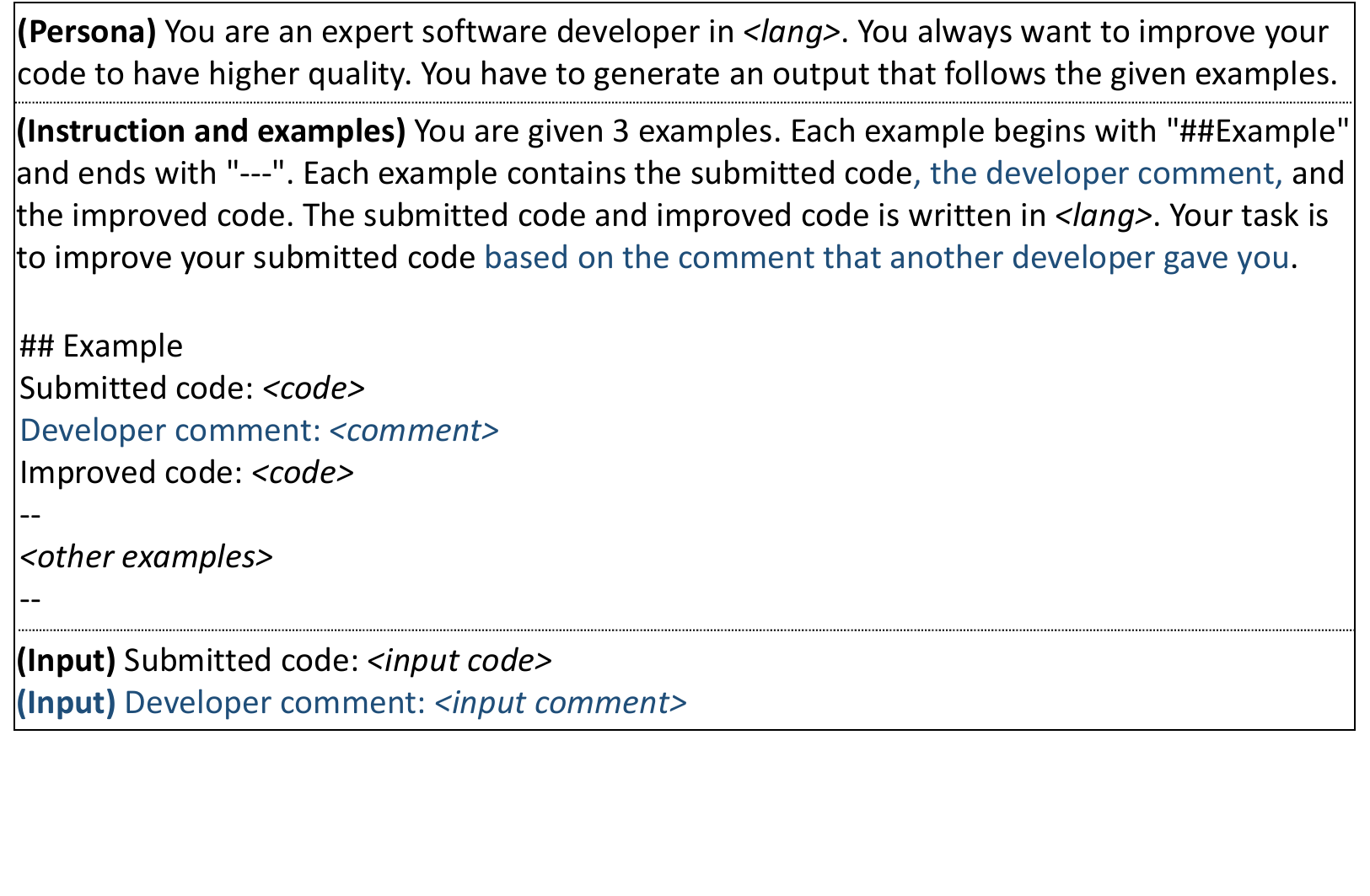}
         \caption{A prompt template for few-shot learning.}
         \label{fig:few-shot prompt}
     \end{subfigure}

    \caption{Prompt templates for zero-shot learning and few-shot learning that contain simple instructions (\textit{lang} refers to a programming language). The text in blue is omitted when reviewers' comments are not used in experiments.}
    \label{fig:prompt-templates}
\end{figure}

\subsection{Inference via Prompting}

In this work, we conduct experiments with the following prompting techniques: zero-shot learning, few-shot learning and a persona.
We explain each prompting technique below.

For \textit{zero-shot learning}, we first design the prompt template as presented in Figure~\ref{fig:zero-shot prompt} by following the guidelines from OpenAI\footref{fn:open-ai-guide-1}\footref{fn:open-ai-guide-2} to ensure that the structure of the prompt is suitable for GPT-3.5.
The prompt template consists of the following components: an instruction and an input (i.e., code submitted for review and a reviewer's comment).

Then, we create prompts by using the prompt template in Figure~\ref{fig:zero-shot prompt} and the code submitted for review with a reviewer's comment in a testing set.
Finally, we use the LLMs to generate revised code from the created prompts.


For \textit{few-shot learning}~\cite{brown2020language, liu2021makes, wei2023larger}, we first design the prompt template as presented in Figure~\ref{fig:few-shot prompt}.
Similar to zero-shot learning, we follow the guidelines from OpenAI when designing the prompt template.
The prompt template consists of the following components: demonstration examples, an instruction and an input (i.e., code submitted for review and a reviewer's comment).


In few-shot learning, demonstration examples are required to create a prompt.
Thus, we select three demonstration examples, where each example consists of two inputs (i.e., code submitted for review and a reviewer's comment) and an output (i.e., revised code), by using BM25~\cite{robertson2009probabilistic}.
We use BM25~\cite{robertson2009probabilistic} since prior work~\cite{gao2023constructing, yuan2023evaluating} shows that BM25~\cite{robertson2009probabilistic} outperforms other sample selection approaches for software engineering tasks.
In this work, we use BM25~\cite{robertson2009probabilistic} provided by the \texttt{gensim}\footnote{https://github.com/piskvorky/gensim} package.
We select three demonstration examples for each testing sample since Gao~\ea~\cite{gao2023makes} showed that GPT-3.5 using three demonstration examples achieves comparable performance (i.e, 90\% of the highest Exact Match) when compared to GPT-3.5 that achieves the highest performance by using 16 or more demonstration examples.

Then, we create prompts from the prompt template in Figure~\ref{fig:few-shot prompt}; the code submitted for review and a reviewer's comment in the testing set; and the demonstration examples of the code submitted for review.
Finally, we use LLMs to generate revised code from the prompts.

For \textit{persona}~\cite{white2023prompt}, we include a persona in the prompt templates in Figure~\ref{fig:prompt-templates} to instruct GPT-3.5 to act as a software developer.
We do so to ensure that the revised code generated by GPT-3.5 looks like the source code written by a software developer.

\begin{table*}[h]
    \centering
    \caption{The evaluation results of GPT-3.5, Magicoder and the existing code review automation approaches.}
    \resizebox{\textwidth}{!}{
    \begin{tabular}{|c|c|c|c|c|c|c|c|c|c|c|c|c|c|c|c|}
    \hline
        \multirow{2}{*}{Approach} & \multirow{2}{*}{Fine-Tuning} & \multicolumn{2}{c|}{Inference Technique} & \multicolumn{2}{c|}{CodeReviewer} & \multicolumn{2}{c|}{Tufano (with comment)} & \multicolumn{2}{c|}{Tufano (without comment)} & \multicolumn{2}{c|}{Android} & \multicolumn{2}{c|}{Google} & \multicolumn{2}{c|}{Ovirt} \\
        \cline{3-16}    
         &    & Prompting & Use Persona & EM & CodeBLEU & EM & CodeBLEU & EM & CodeBLEU & EM & CodeBLEU & EM & CodeBLEU & EM & CodeBLEU \\ \hline
        \multirow{6}{*}{GPT-3.5} & \multirow{2}{*}{\cmark}   & \multirow{4}{*}{Zero-shot} & \xmark    & 37.93\% & 49.00\% & 22.16\% & 82.99\% & 6.02\% & 79.81\% & 2.34\% & 74.15\% & 6.71\% & 81.08\% & 3.05\% & 74.67\% \\
       \cline{4-16}
        ~ &  & ~ & \cmark & 37.70\% & 49.20\% & 21.98\% & 83.04\% & 6.04\% & 79.76\% & 2.29\% & 74.74\% & 6.14\% & 81.02\% & 2.64\% & 74.95\% \\ \cline{2-2}\cline{4-16}
        ~ & \multirow{4}{*}{\xmark} & ~ & \xmark & 17.72\% & 44.17\% & 13.52\% & 78.36\% &  2.62\% & 74.92\% &  0.49\% & 61.85\% & 0.16\% & 61.04\% & 0.48\% & 56.55\% \\ \cline{4-16}
        ~ &  & ~ & \cmark & 17.07\% & 43.11\% & 12.49\% & 77.32\% & 2.29\% & 73.21\% & 0.57\% & 55.88\% & 0.00\% & 50.65\% & 0.22\% & 45.73\% \\ \cline{3-16}
        ~ &  & \multirow{2}{*}{Few-shot} & \xmark & 26.55\% & 47.50\% & 19.79\% & 81.47\% & 8.96\% & 79.21\% & 2.34\% & 75.33\% & 2.89\% & 81.40\% & 1.64\% & 73.83\% \\ \cline{4-16}
        ~ &  & ~ & \cmark & 26.28\% & 47.43\% & 20.03\% & 81.61\% & 9.18\% & 78.98\% & 1.62\% & 74.65\% & 2.45\% & 81.07\% & 1.67\% & 73.29\% \\ 
        \hline
        \multirow{6}{*}{Magicoder} & \multirow{2}{*}{\cmark}   & \multirow{4}{*}{Zero-shot} & \xmark    & 27.43\% & 44.86\% & 11.14\% & 69.77\% & 1.97\% & 69.25\% & 0.27\% & 65.39\% & 0.57\% & 69.30\% & 0.30\% & 64.19\% \\
        \cline{4-16}          &     &       & \cmark   &  27.98\% & 45.36\% & 11.06\% & 69.60\% & 2.12\% & 68.84\% & 0.65\% & 65.41\% & 1.13\% & 69.30\% & 1.00\% & 64.16\% \\
        \cline{2-2}\cline{4-16}          & \multirow{4}{*}{\xmark}    &       & \xmark    & 9.75\% & 39.45\% & 8.65\% & 73.90\% & 0.81\% & 59.49\% & 0.16\% & 47.37\% & 0.08\% & 48.82\% & 0.04\% & 44.38\% \\
       \cline{4-16}          &    &       & \cmark   & 9.93\% & 39.48\% & 8.71\% & 73.57\% & 1.51\% & 67.65\% & 0.27\% & 47.46\% & 0.08\% & 48.58\% & 0.11\% & 43.65\% \\
        \cline{3-16}          &    & \multirow{2}{*}{Few-shot} & \xmark    & 15.89\% & 36.24\% & 2.93\% & 4.36\% & 1.99\% & 7.49\% & 0.22\% & 37.61\% & 0.49\% & 42.50\% & 0.74\% & 40.33\% \\
        \cline{4-16}          &    &       & \cmark   & 17.80\% & 38.93\% & 2.89\% & 3.70\% & 1.84\% & 6.96\% & 0.27\%  & 16.59\% & 0.65\% & 18.83\% & 0.82\% & 19.55\% \\
        \hline
        Guo~\ea~\cite{guo2023exploring} &  \xmark  & Zero-shot &  \cmark  & 21.77\% & 59.85\% & -  & - & -     & - & -  & -  & -  & -  & - & - \\
        \hline
        CodeReviewer~\cite{li2022automating} & \multirow{3}{*}{-} & \multirow{3}{*}{-} & \multirow{3}{*}{-} & 33.23\% & 55.43\% & 15.17\% & 80.83\% & 4.14\% & 78.76\% & 0.54\% & 75.24\% & 0.81\% & 80.10\% & 1.23\% & 75.32\% \\
        \cline{1-1}\cline{5-16}
        TufanoT5~\cite{tufano2022using} & ~ & ~ & ~ & 11.90\% & 43.39\% & 14.26\% & 79.48\% & 5.40\% & 77.26\% & 0.27\% & 75.88\% & 1.37\% & 82.25\% & 0.19\% & 73.53\% \\
        \cline{1-1}\cline{5-16}
        D-ACT~\cite{pornprasit2023d} & ~ & ~ & ~ & - & - & - & - & - & - & 0.65\% & 75.99\% & 5.98\% & 81.85\% & 1.79\% & 79.77\% \\ \hline
    \end{tabular}
    }
    \label{tab:result}%
\end{table*}

\subsection{The Evaluation Measures}
We use the following measures to evaluate the performance of the studied LLMs (i.e., GPT-3.5 and Magicoder~\cite{wei2023magicoder}) and code review automation approaches (i.e., CodeReviewer~\cite{li2022automating}, TufanoT5~\cite{tufano2022using}, and D-ACT~\cite{pornprasit2023d}).

    \begin{enumerate}
        \item \textbf{Exact Match (EM)}~\cite{pornprasit2023d, li2022automating, tufano2022using} is the number of the generated revised code that is the same as the actual revised code in the testing dataset.
        We use this measure since it is widely used for evaluating code review automation approaches~\cite{tufano2019learning, pornprasit2023d, tufano2022using}.
        To compare the generated revised code with the actual revised code, we first tokenize both revised code to sequences of tokens.
        Then, we compared the sequence of tokens of the generated revised code with the sequence of tokens of the actual revised code.
        A high value of EM indicates that a model can generate revised code that is the same as the actual revised code in the testing dataset.
        
        \item \textbf{CodeBLEU}~\cite{ren2020codebleu} is the extended version of BLEU (i.e., an n-gram overlap between the translation generated by a deep learning model and the translation in ground tr\-uth)~\cite{bleu} for automatic evaluation of the generated code.
        We do not measure BLEU like in prior work~\cite{li2022automating, tufano2022using} since Ren~\ea~\cite{ren2020codebleu} found that this measure ignores syntactic and semantic correctness of the generated code.
        In addition to BLEU, CodeBLEU considers the weighted n-gram match, matched syntactic information (i.e., abstract syntax tree: AST) and matched semantic information (i.e., data flow: DF) when computing the similarity between the generated revised code and the actual revised code.
        A high value of CodeBLEU indicates that a model can generate revised code that is syntactically and semantically similar to the actual revised code in the testing dataset.

    \end{enumerate}

\subsection{The Hyper-Parameter Settings}

    In this study, we use the following hyper-parameter settings when using GPT-3.5 to generate revised code: temperature of 0.0 (as suggested by Guo~\ea~\cite{guo2023exploring}), top\_p of 1.0 (default value), and max length of 512.
    To fine-tune GPT-3.5, we use hyper-parameters  (e.g., number of epochs and learning rate) that are automatically provided by OpenAI API.

    For Magicoder~\cite{wei2023magicoder}, we use the same hyper-parameter as GPT-3.5 to generate revised code.
    To fine-tune Magicoder, we use the following hyper-parameters for DoRA~\cite{liu2024dora}: attention dimension ($r$) of 16, alpha ($\alpha$) of 8, and dropout of 0.1 .

\begin{table}[ht]
  \centering
  \caption{The statistical details of GPT-3.5, Magicoder and the existing code review automation approaches.}
    \begin{tabular}{|c|c|}
    \hline
    Model & \# parameters \\
    \hline
    GPT-3.5 & 175 B \\
    \hline
    Magicoder~\cite{wei2023magicoder} & 6.7 B \\ 
    \hline
    TufanoT5~\cite{tufano2022using} & 60.5 M \\
    \hline
    CodeReviewer~\cite{li2022automating} & 222.8 M \\
    \hline
    D-ACT~\cite{pornprasit2023d} & 222.8 M \\
    \hline
    \end{tabular}%
  \label{tab:model-detail}%
\end{table}%

\section{Result} \label{sec:results}

In this section, we present the
results of the following three research questions.

\noindent \textbf{(RQ1) \rqone}

\noindent \smallsection{Approach}
To address this RQ, we leverage fine-tuning and inference techniques (i.e., zero-shot learning, few-shot learning, and persona) on GPT-3.5 and Magicoder \revised{R2-5}{
(The details of GPT-3.5 and Magicoder are presented in Table~\ref{tab:model-detail})
}.
Then, we measure EM of the results obtained from GPT-3.5, Magicoder and Guo~\ea's approach~\cite{guo2023exploring}.

\begin{figure*}[h]
     \centering
     \begin{subfigure}[]{\columnwidth}
         \centering
         \includegraphics[width=\columnwidth, trim = {0 0 0 0}, clip, page=1]{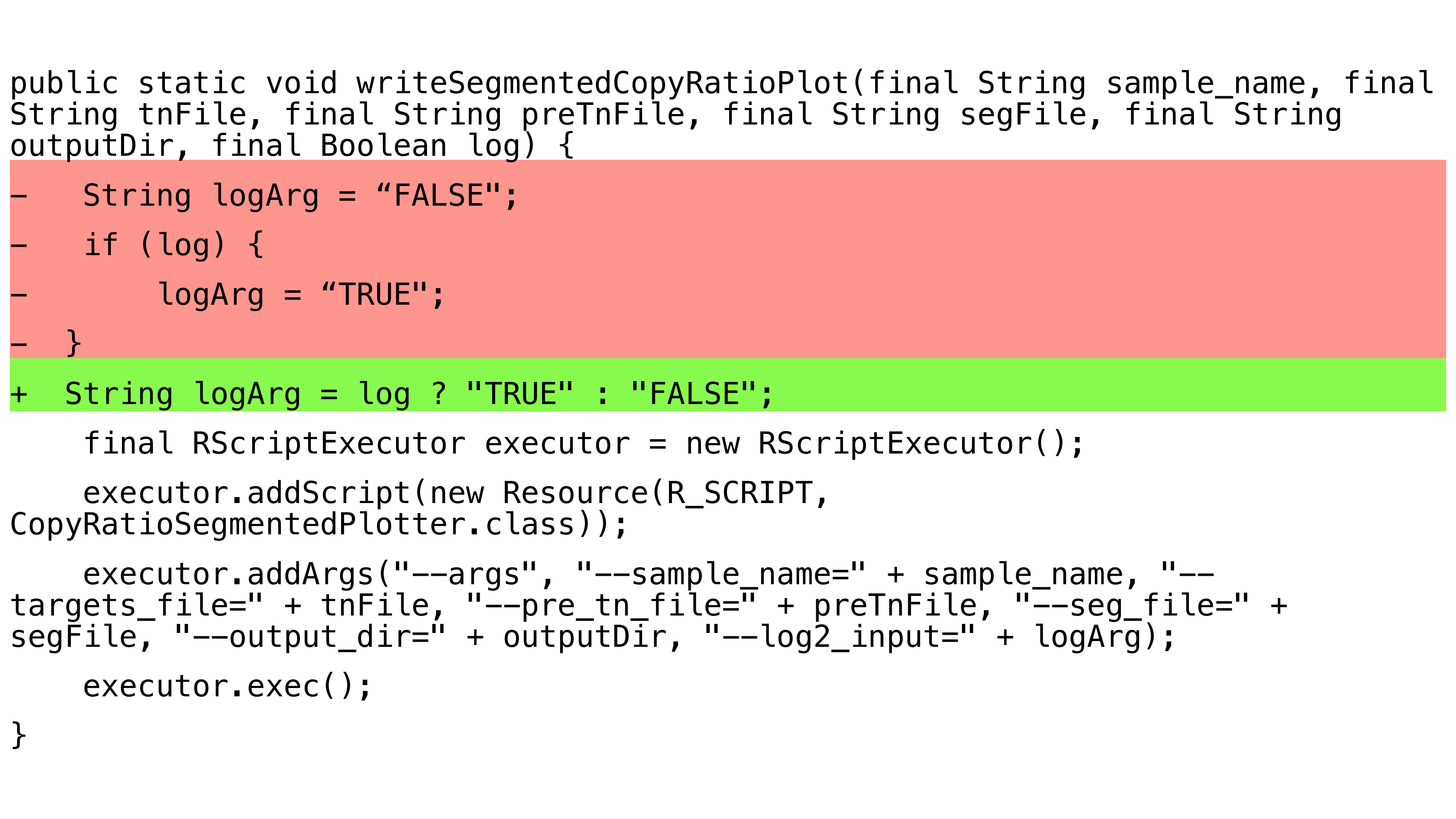}
         \caption{The difference between code submitted for review and revised code that GPT-3.5 with zero-shot learning (no persona) correctly generates.}
         \label{fig:RQ3-0-shot-no-persona}
     \end{subfigure}
     \begin{subfigure}[]{\columnwidth}
         \centering
         \includegraphics[width=\columnwidth, trim = {0
         0 0 0}, clip, page=2]{figure/sample-diff-with-and-without-persona.pdf}
         \caption{The difference between code submitted for review and revised code that GPT-3.5 with zero-shot learning (use persona) incorrectly generates.}
         \label{fig:RQ3-0-shot-with-persona}
     \end{subfigure}
     \par\bigskip 
     \begin{subfigure}[]{\columnwidth}
         \centering
         \includegraphics[width=\columnwidth, trim = {0 0 0 0}, clip, page=3]{figure/sample-diff-with-and-without-persona.pdf}
         \caption{The difference between code submitted for review and revised code that GPT-3.5 with few-shot learning (no persona) correctly generates.}
        \label{fig:RQ3-few-shot-no-persona}
     \end{subfigure}
    \begin{subfigure}[]{\columnwidth}
         \centering
        \includegraphics[width=\columnwidth, trim = {0 0 0 0}, clip, page=4]{figure/sample-diff-with-and-without-persona.pdf}
         \caption{The difference between code submitted for review and revised code that GPT-3.5 with few-shot learning (use persona) incorrectly generates.}
         \label{fig:RQ3-few-shot-with-persona}
     \end{subfigure}
     
    \caption{(RQ3) Examples of the difference between code submitted for review and revised code generated by GPT-3.5 with zero-shot learning and few-shot learning.}
    \label{fig:RQ3-code-change-with-and-without-persona}
\end{figure*}

\noindent \smallsection{Result}
\textbf{The fine-tuning of GPT 3.5 with zero-shot learning helps GPT-3.5 to achieve 73.17\% -74.23\% higher EM than the Guo~\ea~\cite{guo2023exploring}'s approach.}
Table~\ref{tab:result} shows the results of EM achieved by GPT-3.5, Magicoder and Guo~\ea's approach~\cite{guo2023exploring}.
The table shows that when GPT-3.5 and Magicoder are fine-tuned, such models achieve 73.17\% -74.23\% and 26.00\% - 28.53\% higher EM than the Guo~\ea's approach~\cite{guo2023exploring}, respectively.

The results indicate that model fine-tuning could help GPT-3.5 to achieve higher EM when compared to the Guo~\ea's approach~\cite{guo2023exploring}.
The higher EM has to do with model fine-tuning.
When GPT-3.5 or Magicoder is fine-tuned, such models learn the relationship between inputs (i.e., code submitted for review and a reviewer’s comment) and an output (i.e., revised code) from a number of examples in a training set.
On the contrary, Guo~\ea's approach~\cite{guo2023exploring} only relies on the instruction and given input to generate revised code, which GPT-3.5 never learned during model pre-training.



\noindent \textbf{(RQ2) \rqtwo}

\noindent \smallsection{Approach}
To address this RQ, we fine-tune GPT-3.5 as explained in Section~\ref{sec:study-design}.
Then, we measure EM and CodeBLEU of the results obtained from the fine-tuned GPT-3.5 and the non fine-tuned GPT-3.5 with zero-shot learning. 


\noindent \smallsection{Result}
\textbf{The fine-tuning of GPT 3.5 with zero-shot learning helps GPT-3.5 to achieve 63.91\% - 1,100\% higher EM than those that are not fine-tuned.}
Table~\ref{tab:result} shows that in terms of EM, the fine-tuning of GPT 3.5 with zero-shot learning helps GPT-3.5 to achieve 63.91\% - 1,100\% higher than those that are not fine-tuned.
In terms of CodeBLEU, the fine-tuning of GPT 3.5 with zero-shot learning helps GPT-3.5 to achieve 5.91\% - 63.9\% higher than those that are not fine-tuned.

The results indicate that fine-tuned GPT-3.5 achieve higher EM and CodeBLEU than those that are not fine-tuned.
During the model fine-tuning process, GPT-3.5 adapt to the code review automation task by directly learning the relationship between inputs (i.e., code submitted for review and a reviewer's comment) and an output (i.e., revised code) from a number of examples in a training set.
In contrast, non fine-tuned GPT-3.5 is given only an instruction and inputs which are not presented during model pre-training.
Therefore, fine-tuned GPT-3.5 can better adapt to the code review automation task than those that are not fine-tuned.

\begin{figure*}[h]
     \centering
     \begin{subfigure}[]{\columnwidth}
         \centering
         \includegraphics[width=\columnwidth, trim = {1 25cm 0 0}, clip, page=1]{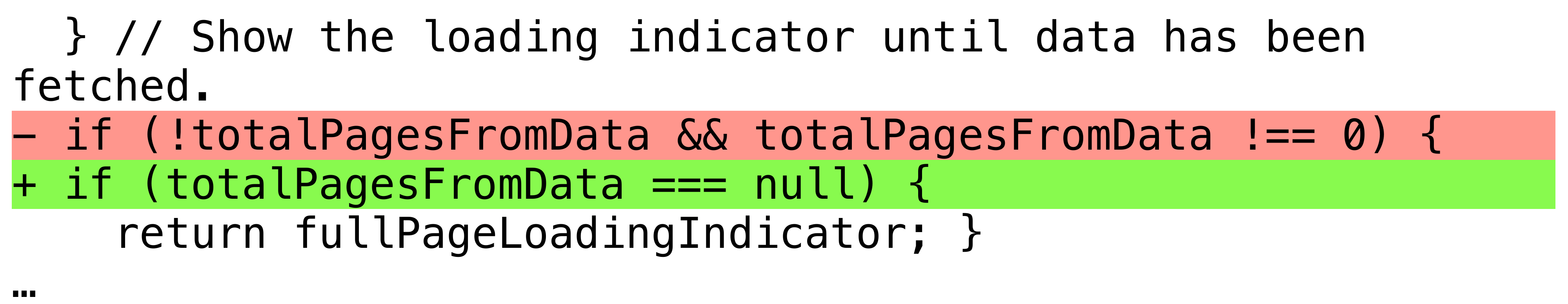}
         \caption{Example of the code change for \textit{bug fixing} (modify if condition)}
         \label{fig:RQ3-bug-fix-1}
     \end{subfigure}
     \begin{subfigure}[]{\columnwidth}
         \centering
         \includegraphics[width=\columnwidth, trim = {0 26cm
         0 0 0}, clip, page=2]{figure/sample-code-change-each-type.pdf}
         \caption{Example of the code change for \textit{bug fixing} (remove \texttt{synchronized} keyword)}
         \label{fig:RQ3-bug-fix-2}
     \end{subfigure}
     \par\bigskip 
     \begin{subfigure}[]{\columnwidth}
         \centering
         \includegraphics[width=\columnwidth, trim = {0 25cm 0 0}, clip, page=3]{figure/sample-code-change-each-type.pdf}
         \caption{Example of the code change for \textit{refactoring} (change variable name)}
         \label{fig:RQ3-refactor-1}
     \end{subfigure}
    \begin{subfigure}[]{\columnwidth}
         \centering
        \includegraphics[width=\columnwidth, trim = {0 25cm 0 0}, clip, page=4]{figure/sample-code-change-each-type.pdf}
         \caption{Example of the code change for \textit{refactoring} (remove \texttt{this} qualifier)}
         \label{fig:RQ3-refactor-2}
     \end{subfigure}
     \par\bigskip 
     \begin{subfigure}[]{\columnwidth}
         \centering
         \includegraphics[width=\columnwidth, trim = {0 15cm 0 0}, clip, page=5]{figure/sample-code-change-each-type.pdf}
         \caption{Example of the code change for \textit{other} (remove if condition)}
         \label{fig:RQ3-other-1}
     \end{subfigure}
    \begin{subfigure}[]{\columnwidth}
         \centering
        \includegraphics[width=\columnwidth, trim = {0 15cm 0 0}, clip, page=6]{figure/sample-code-change-each-type.pdf}
         \caption{Example of the code change for \textit{other} (remove method call)}
         \label{fig:RQ3-other-2}
     \end{subfigure}
     
    \caption{Example of the code changes of each type}
    \label{fig:RQ3-code-change-example}
\end{figure*}

\noindent \textbf{(RQ3) \rqthree}

\noindent \smallsection{Approach}
To address this RQ, we use zero-shot learning and few-shot learning with \textit{non fine-tuned} GPT-3.5,  where each inference technique is used with and without a persona, to generate revised code as explained in Section~\ref{sec:study-design}.
Then, similar to RQ2, we measure EM and CodeBLEU of the results obtained from GPT-3.5.

\noindent \smallsection{Result}
\textbf{GPT-3.5 with few-shot learning achieves 46.38\% - 659.09\% higher EM than GPT-3.5 with zero-shot learning.}
Table~\ref{tab:result} shows that in terms of EM, the use of few-shot learning on GPT-3.5 helps GPT-3.5 to achieve 46.38\% - 241.98\% and 53.95\% - 659.09\% higher than the use of zero-shot learning on GPT-3.5 without a persona and with a persona, respectively.
In terms of CodeBLEU, the use of few-shot learning on GPT-3.5 helps GPT-3.5 to achieve  3.97\% - 33.36\% and 5.55\% - 60.27\% higher than the use of zero-shot learning on GPT-3.5 without a persona and with a persona, respectively.

The results indicate that GPT-3.5 with few-shot learning can achieve higher EM and CodeBLEU than GPT-3.5 with zero-shot learning.
When few-shot learning is used to generate revised code from GPT-3.5, GPT-3.5 has more information from given demonstration examples in a prompt to guide the generation of revised code from given code submitted for review and a reviewer's comment.
In other words, such demonstration examples could help GPT-3.5 to correctly generate revised code from a given code submitted for review and a reviewer's comment.

\textbf{When a persona is included in input prompts, GPT-3.5 achieves 1.02\% - 54.17\% lower EM than when the persona is not included in input prompts.}
Table~\ref{tab:result} also shows that when a persona is included in prompts, GPT-3.5 with zero-shot and few-shot learning
achieves 3.67\% - 54.17\% and 1.02\% - 30.77\% lower EM compared to when a persona is not included in prompts, respectively.
Similarly, when a persona is included in prompts, GPT-3.5 with zero-shot and few-shot learning achieves 1.33\% - 19.13\% and 0.15\% - 0.90\% lower CodeBLEU compared to when a persona is not included in prompts, respectively. 


The results indicate that when a persona is included in prompts, GPT-3.5 with zero-shot and few-shot learning achieves lower EM and CodeBLEU.
\revised{R2-2}{
To illustrate the impact of the persona, we present an example of the revised code that GPT-3.5 generates when zero-shot learning with and without a persona is used, and an example of the revised code that GPT-3.5 generates when few-shot learning with and without a persona is used in Figure~\ref{fig:RQ3-code-change-with-and-without-persona}.

Figure~\ref{fig:RQ3-0-shot-no-persona} presents the revised code that GPT-3.5 with zero-shot learning (no persona)  correctly generates.
In this figure, GPT-3.5 suggests an alternative way to initialize variable \texttt{logArg}.
In contrast, Figure~\ref{fig:RQ3-0-shot-with-persona} presents the revised code that GPT-3.5 with zero-shot learning (use persona) incorrectly generates.
In this figure, GPT-3.5 suggests changing the variable type from \texttt{Boolean} to \texttt{boolean} and changing the variable name from \texttt{sample\_name} to \texttt{sampleName} in addition to suggesting an alternative way to initialize variable \texttt{logArg}.

Figure~\ref{fig:RQ3-few-shot-no-persona} presents another example of the revised code that GPT-3.5 with few-shot learning (no persona)  correctly generates.
In this figure, GPT-3.5 suggests a new \texttt{if} condition.
On the contrary, Figure~\ref{fig:RQ3-few-shot-with-persona} presents the revised code that GPT-3.5 with few-shot learning (use persona) incorrectly generates.
In this figure, GPT-3.5 suggests an additional \texttt{if} statement and an additional \texttt{else} block.

The above examples imply that when a persona is included in prompts, GPT-3.5 tends to suggest additional incorrect changes to the submitted code compared to when a persona is not included in prompts.
}


The above results indicate that the best prompting strategy when using GPT-3.5 without fine-tuning is few-shot learning without a persona.

\section{Discussion} \label{sec:discussion}

In this section, we discuss the implications of our findings, the additional results of GPT-3.5, and the cost and benefits of using GPT-3.5.

\begin{figure*}[h]
     \centering
     \begin{subfigure}[]{\textwidth}
         \centering
         \includegraphics[width=\textwidth, trim = {0.5cm 0.7cm 0 0}, clip]{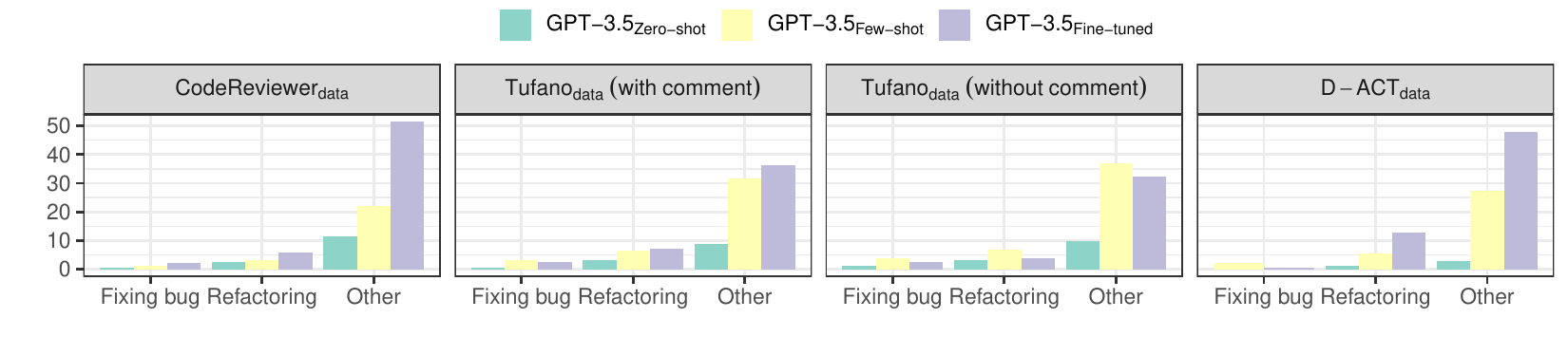}
         \caption{The results of GPT-3.5 (persona is included in prompts)}
         \label{fig:RQ3-analysis-with-persona}
     \end{subfigure}
     \par\bigskip 
     \begin{subfigure}[]{\textwidth}
         \centering
         \includegraphics[width=\textwidth, trim = {0.5cm 0.7cm 0 0}, clip]{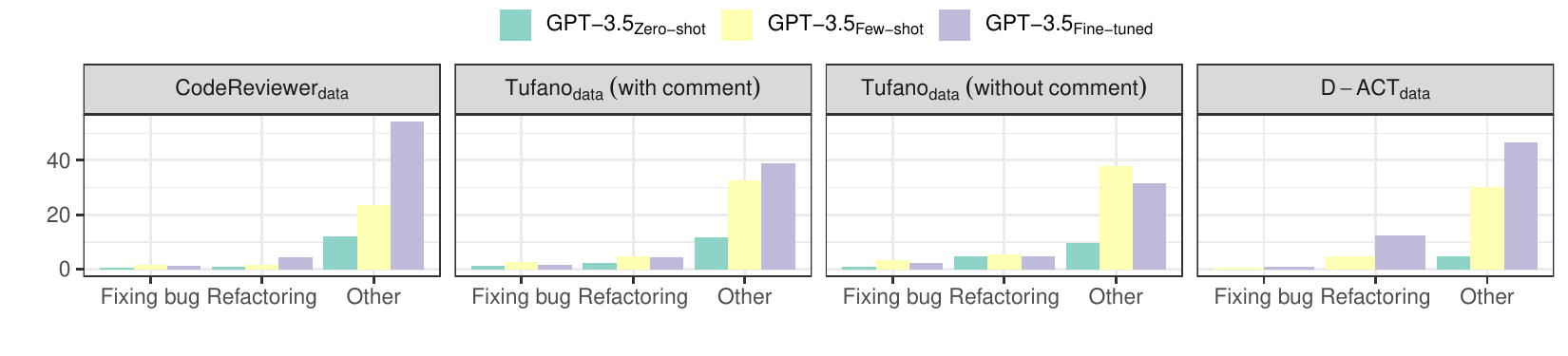}
      \caption{The results of GPT-3.5 (persona is not included in prompts)}
     \label{fig:RQ3-analysis-without-persona}
     \end{subfigure}

    \caption{The EM achieved by \ChatGPTZeroShot, \ChatGPTFewShot~and \ChatGPTFineTune~categorized by the types of code change. 
    Here, \ChatGPTZeroShot~and \ChatGPTFewShot~refer to non fine-tuned GPT-3.5 with zero-shot learning and few-shot learning, respectively.
    On the other hand, GPT-3.5\textsubscript{Fine-tuned} refers to fine-tuned GPT-3.5 with zero-shot learning.}
    \label{fig:RQ3-analysis}
\end{figure*}

\subsection{Implications of Our Findings}

\revised{R1-4}{
\textbf{GPT-3.5 does not require a lot of training data for model fine-tuning to adapt to the code review automation task} since Table~\ref{tab:result} shows that GPT-3.5 that is fine-tuned on a subset of a training set outperforms the studied code review automation approaches~\cite{pornprasit2023d, li2022automating, tufano2022using}.
The results imply that GPT-3.5 can adapt to the code review automation task by learning from a small set of training examples (approximately 20k training examples in this study), unlike the studied code review automation approaches that require the whole training set to adapt to the code review automation task. 

}

\textbf{Recommendations to practitioners.}
LLMs for code review automation should be fine-tuned to achieve the highest performance.
The reason for this recommendation is the results of RQ2 show that fine-tuned GPT-3.5 outperforms those that are not fine-tuned.
In contrast, when data is not sufficient for model fine-tuning (e.g., a cold-start problem), few-shot learning without a persona should be used for LLMs for code review automation.
The reason for this recommendation is the results of RQ3 show that GPT-3.5 with few-shot learning outperforms GPT-3.5 with zero-shot learning, and GPT-3.5 without a persona outperforms GPT-3.5 with persona.

\subsection{The Characteristics of the Revised Code that are Correctly Generated by GPT-3.5}

The results of RQ2 and RQ3 demonstrate the benefits of model fine-tuning and few-shot learning on GPT-3.5 for the code review automation task, respectively.
However, practitioners still do not clearly understand the characteristics of the code changes of the revised code that GPT-3.5 correctly generates.
To address this challenge, we aim to qualitatively investigate the revised code that GPT-3.5 can correctly generate.
To do so, we randomly select the revised code that is only correctly generated by a particular model (e.g., we randomly obtain the revised code that only fine-tuned GPT-3.5 correctly generates while the others do not.) by using a confidence level of 95\% and a confidence interval of 5\%.
\revised{R2-3}{
Then, we classify the code changes of the selected revised code into the following categories based on the taxonomy of code change created by Tufano~\ea~\cite{tufano2019learning} (the examples of the code changes in each category are depicted in Figure~\ref{fig:RQ3-code-change-example}):

\begin{itemize}
    \item \textit{fixing bug}: The code changes in this category involve fixing bugs in the past.
    The following sub-categories are related to this category: exception handling, conditional statement, lock mechanism, method return value, and me\-thod invocation.
    \item \textit{refactoring}: The code changes in this category involve making changes to code structure without changing the behavior of the changed code.
    The following sub-categories are related to this category: inheritance; encapsulation; methods interaction; readability; and renaming parameter, method and variable.
    \item \textit{other}: The code changes that cannot be classified as neither \textit{fixing bug} nor \textit{refactoring} will fall into this category.
\end{itemize}
}

Figure~\ref{fig:RQ3-analysis} presents the characteristics of the code changes (i.e., \textit{Fixing Bug}, \textit{Refactoring} and \textit{Other}) of the revised code that GPT-3.5 correctly generates.
According to the figure, for the Tufano\textsubscript{data} (with comment), CodeReviewer\textsubscript{data} and D-ACT\textsubscript{data} dataset, the fine-tuning of GPT 3.5 with zero-shot learning helps GPT-3.5 achieves the highest EM for the code changes of \textit{Refactoring} and \textit{Other}.
In contrast, for the Tufano\textsubscript{data} (without comment) dataset, non fine-tuned GPT-3.5 with few-shot learning achieves the highest EM for the code changes of all categories.
\revised{R2-3}{
According to Figure~\ref{fig:RQ3-analysis-with-persona} and Figure~\ref{fig:RQ3-analysis-without-persona}, we also find that fine-tuned GPT-3.5 and non fine-tuned GPT-3.5 that zero-shot and few-shot learning are used achieve the highest EM for the code changes of type \textit{other} across all studied datasets.
The reason for this result is that we do not specify the characteristics of the revised code (i.e., \textit{fixing bug} and \textit{refactoring}) in prompts.
Thus, such models possibly generate revised code that is not specific to \textit{fixing bug} or \textit{refactoring}.
Therefore, the majority of the code changes of the generated revised code are categorized as \textit{other}.
}

\begin{table*}[h]
  \centering
  \caption{The evaluation results of GPT-3.5 when being fine-tuned with different sizes of training sets.}
  \resizebox{\linewidth}{!}{
    \begin{tabular}{|c|c|c|c|c|c|c|c|c|c|c|c|c|}
    \cline{2-13}    \multicolumn{1}{c|}{}
      & \multicolumn{2}{c|}{CodeReviewer} & \multicolumn{2}{c|}{Tufano (with comment)} & \multicolumn{2}{c|}{Tufano (without comment)} & \multicolumn{2}{c|}{Android} & \multicolumn{2}{c|}{Google} & \multicolumn{2}{c|}{Ovirt}  \\
\hline \makecell{Size of \\Training Set}       & EM    & CodeBLEU & EM    & CodeBLEU & EM    & CodeBLEU & EM    & CodeBLEU & EM    & CodeBLEU & EM    & CodeBLEU \\
    \hline
    6\%   & 37.93\% & 49.00\% & 22.16\% & 82.99\% & \textbf{6.02\%} & 79.81\% & 2.34\% & 74.15\% & 6.71\% & 81.08\% & \textbf{3.05\%} & 74.67\% \\
    \hline
    10\%  & 37.72\% & 48.83\% & 22.31\% & 83.43\% & 5.37\% & \textbf{80.68\%} & \textbf{2.51\%} & 75.10\% & 5.98\% & 80.65\% & 2.71\% & 75.13\% \\
    \hline
    20\%  & \textbf{38.80\%} & \textbf{49.33\%} & \textbf{22.84\%} & \textbf{83.44\%} & 5.65\% & 80.42\% & 2.34\% & \textbf{76.04\%} & \textbf{7.52\%} & \textbf{81.40\%} & 2.83\% & \textbf{75.46\%} \\
    \hline
    \end{tabular}%
  }
  \label{tab:result-more-data}%
\end{table*}%

\revisedTwo{R1-2}
{
\subsection{The Impact of the Size of Training Dataset on Fine-Tuned GPT-3.5}

The results of RQ2 show that model fine-tuning can help increase the performance of GPT-3.5.
However, little is known whether fine-tuned GPT-3.5 can achieve higher performance when being fine-tuned with larger training sets.
Therefore, we conduct experiments by using 10\% and 20\% of training sets to fine-tune GPT-3.5 (we do not use persona in these experiments).

Table~\ref{tab:result-more-data} shows the results of EM and CodeBLEU that fine-tuned GPT-3.5 achieves across different sizes of training sets.
The table shows that GPT-3.5 that is fine-tuned with 20\% of a training set achieves 2.29\% - 12.07\% higher EM and 0.54\% - 2.55\% higher CodeBLEU than GPT-3.5 that is fine-tuned with 6\% of a training set.
In addition, GPT-3.5 that is fine-tuned with 20\% of a training set achieves 2.38\% - 25.75\% higher EM and 0.01\% - 1.25\% higher CodeBLEU than GPT-3.5 that is fine-tuned with 10\% of a training set, respectively.
The results indicate that fine-tuned GPT-3.5 achieves higher performance when being fine-tuned with a larger training set.
}

\begin{figure}[h]
    \centering
    \begin{subfigure}{\columnwidth}
         \centering
         \includegraphics[width=\columnwidth, page = 1, trim = {0 3.5cm 0 0}, clip]{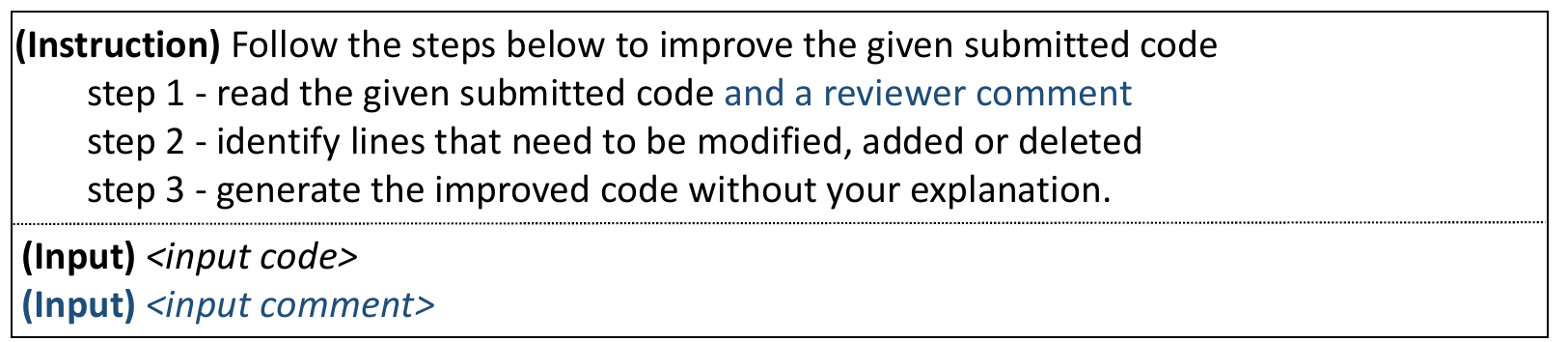}
         \caption{A prompt template for zero-shot learning.}
         \label{fig:zero-shot prompt-task-decomposition}
     \end{subfigure}
     \par\bigskip
     \begin{subfigure}{\columnwidth}
         \centering
         \includegraphics[width=\columnwidth, page = 1, trim = {0 0 0 0}, clip]{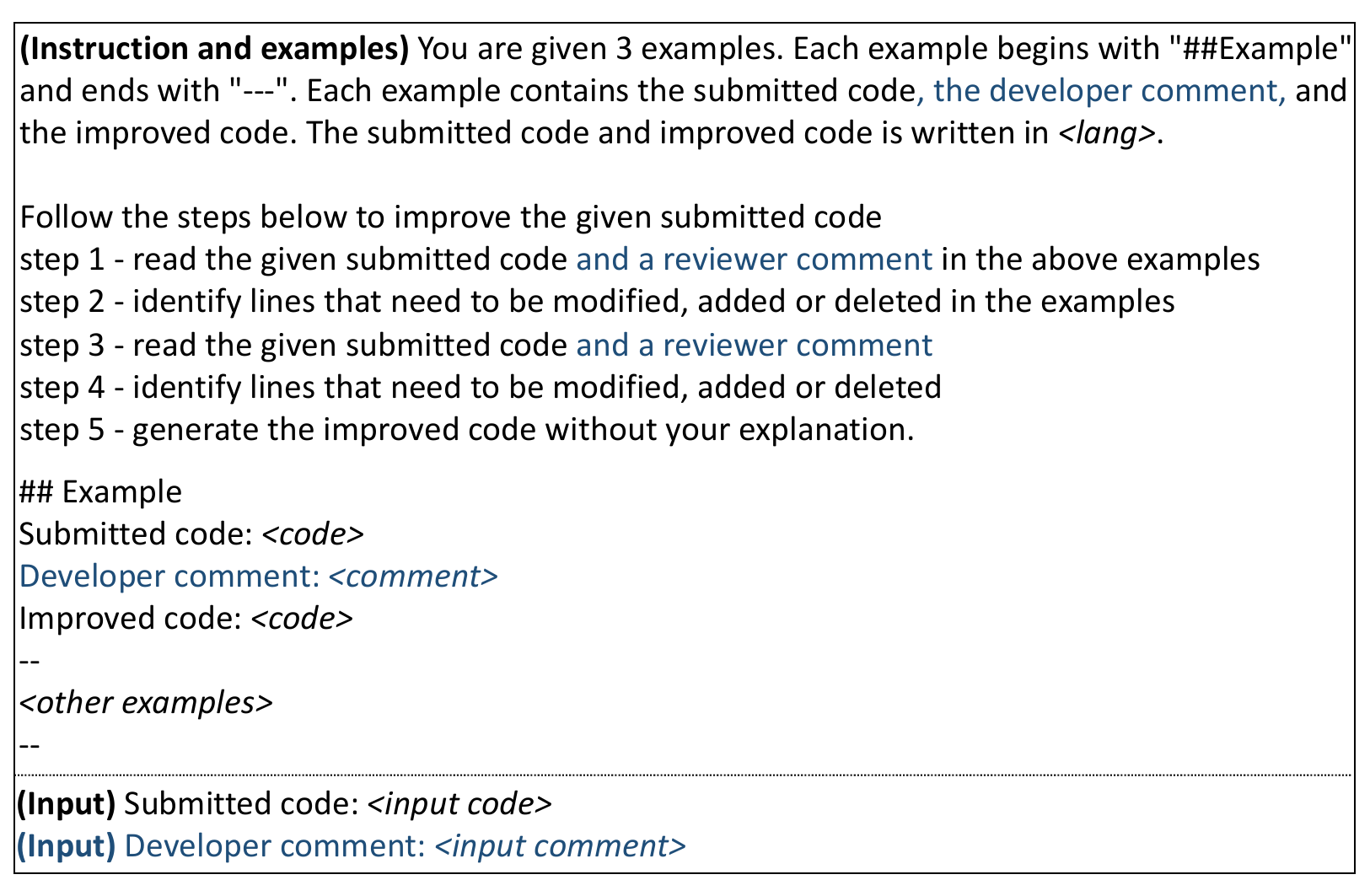}
         \caption{A prompt template for few-shot learning.}
         \label{fig:few-shot prompt-task-decomposition}
     \end{subfigure}

    \caption{Prompt templates for zero-shot learning and few-shot learning that the instructions are broken into smaller tasks (\textit{lang} refers to a programming language). The text in blue is omitted when reviewers' comments are not used in experiments.}
    \label{fig:prompt-templates-task-decomposition}
\end{figure}

\begin{figure}[h]
    \centering
    \begin{subfigure}{\columnwidth}
         \centering
         \includegraphics[width=\columnwidth, page = 1, trim = {0 2cm 0 0}, clip]{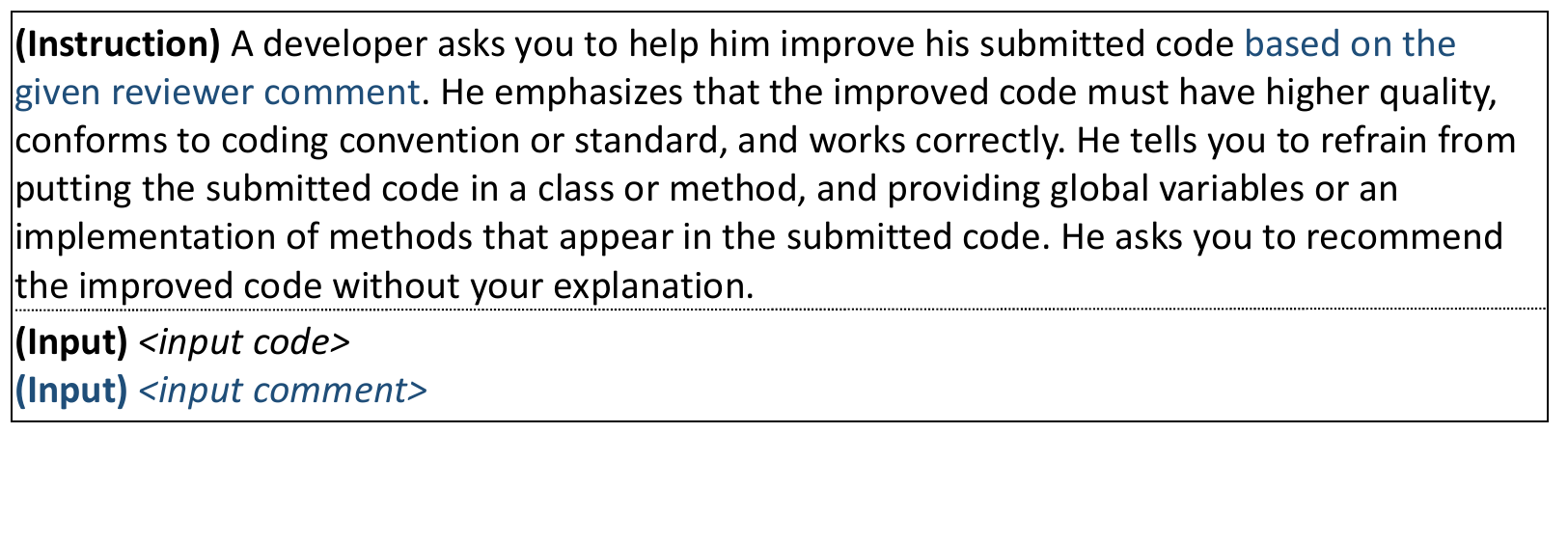}
         \caption{A prompt template for zero-shot learning.}
         \label{fig:zero-shot prompt-more-detail}
     \end{subfigure}
     \par\bigskip
     \begin{subfigure}{\columnwidth}
         \centering
         \includegraphics[width=\columnwidth, page = 1, trim = {0 0 0 0}, clip]{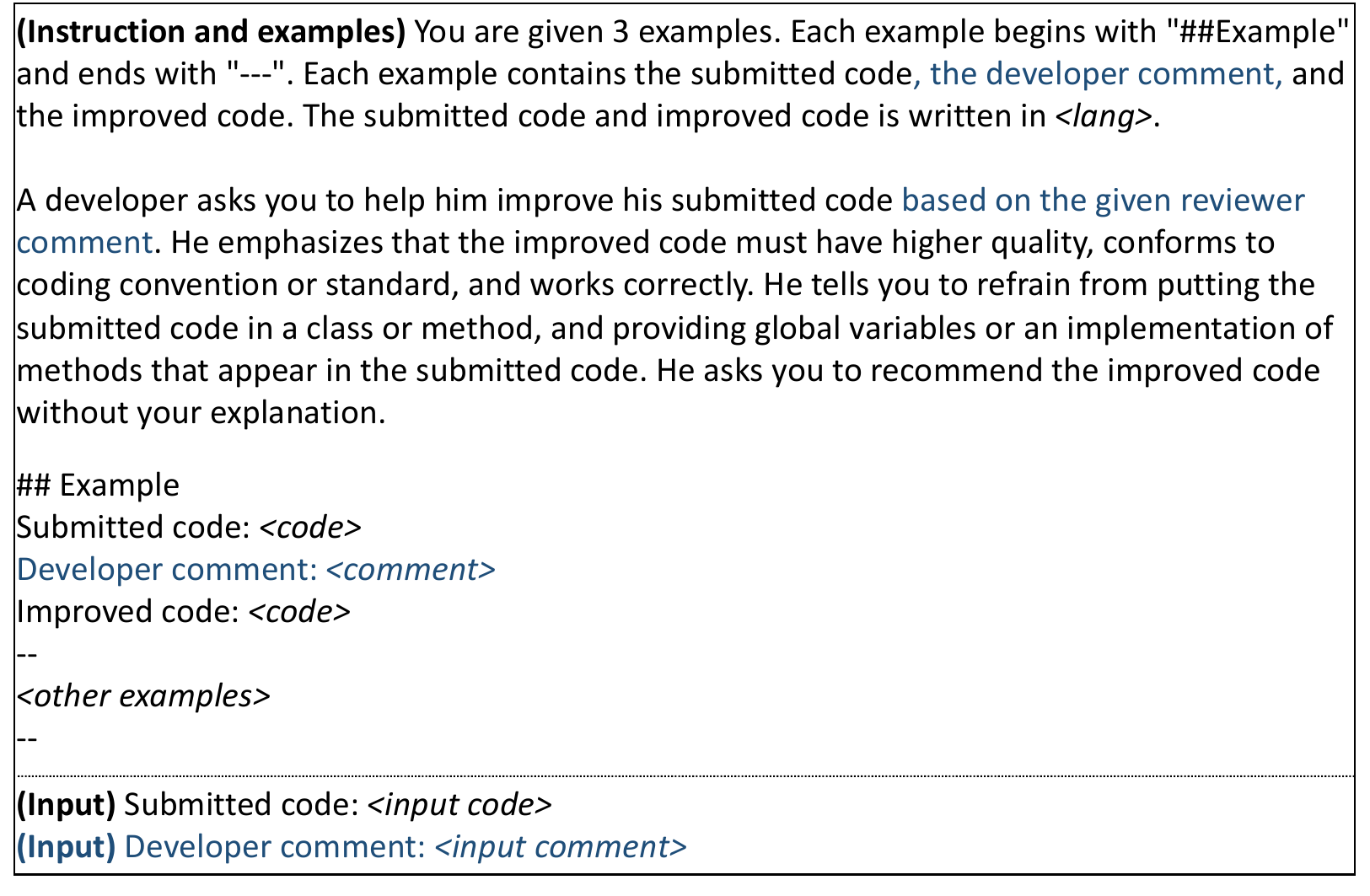}
         \caption{A prompt template for few-shot learning.}
         \label{fig:few-shot prompt-more-detail}
     \end{subfigure}

    \caption{Prompt templates for zero-shot learning and few-shot learning that more details are added to the instructions (\textit{lang} refers to a programming language). The text in blue is omitted when reviewers' comments are not used in experiments.}
    \label{fig:prompt-templates-more-detail}
\end{figure}

\begin{table*}[htbp]
  \centering
  \caption{The evaluation results of GPT-3.5 for different prompt templates. 
  P1 refers to the prompt templates with simple instructions (Figure~\ref{fig:prompt-templates}).
  P2 refers to the prompt templates with instructions being broken down into smaller steps.(Figure~\ref{fig:prompt-templates-task-decomposition}).
  P3 refers to the prompt templates with detailed instructions (Figure~\ref{fig:prompt-templates-more-detail}).}
  \resizebox{\linewidth}{!}{
    \begin{tabular}{|c|c|c|c|c|c|c|c|c|c|c|c|c|c|}
\cline{3-14}    \multicolumn{1}{r}{} &       & \multicolumn{2}{c|}{CodeReviewer} & \multicolumn{2}{c|}{Tufano (with comment)} & \multicolumn{2}{c|}{Tufano (without comment)} & \multicolumn{2}{c|}{Android} & \multicolumn{2}{c|}{Google} & \multicolumn{2}{c|}{Ovirt} \\
    \hline
    \makecell{Prompt\\Design} & Prompting & EM    & CodeBLEU & EM    & CodeBLEU & EM    & CodeBLEU & EM    & CodeBLEU & EM    & CodeBLEU & EM    & CodeBLEU \\
    \hline
    P1    & \multirow{3}{*}{Zero-shot} & 17.72\% & 44.17\% & 13.52\% & 78.36\% & 2.62\% & 74.92\% & 0.49\% & 61.85\% & 0.16\% & 61.04\% & 0.48\% & 56.55\% \\
\cline{1-1}\cline{3-14}    P2    &       & 14.47\% & 43.52\% & 11.24\% & 79.05\% & 2.25\% & 76.54\% & 0.54\% & 66.10\% & 0.16\% & 67.07\% & 0.33\% & 60.76\% \\
\cline{1-1}\cline{3-14}    P3    &       & 11.94\% & 41.18\% & 9.86\% & 76.18\% & 1.26\% & 72.31\% & 0.05\% & 53.03\% & 0.08\% & 46.22\% & 0.26\% & 41.20\% \\
    \hline
    P1    & \multirow{3}{*}{Few-shot} & 26.55\% & 47.50\% & 19.79\% & 81.47\% & 8.96\% & 79.21\% & 2.34\% & 75.33\% & 2.89\% & 81.40\% & 1.64\% & 73.83\% \\
\cline{1-1}\cline{3-14}    P2    &       & 25.25\% & 48.45\% & 15.82\% & 80.16\% & 6.84\% & 76.50\% & 0.60\% & 75.94\% & 3.56\% & 81.40\% & 1.67\% & 74.18\% \\
\cline{1-1}\cline{3-14}    P3    &       & 25.14\% & 48.60\% & 15.07\% & 79.83\% & 5.81\% & 75.76\% & 0.38\% & 74.95\% & 2.91\% & 81.18\% & 1.49\% & 73.31\% \\
    \hline
    \end{tabular}%
  }
  \label{tab:result-multi-prompt}%
\end{table*}%

\revisedTwo{R1-4}
{
\subsection{The Impact of Prompt Design on GPT-3.5}

In RQ3, we use the prompt templates in Figure~\ref{fig:prompt-templates} that contain simple instructions to conduct experiments.
However, prior work~\cite{obrien2024prompt, hossain2024deep} found that the design of prompts has an impact on the performance of LLMs.
Thus, we further investigate the impact of prompt design on GPT-3.5 for code review automation.
To do so, we conduct experiments by using the following two new prompt designs (we do not include a persona in these prompt designs since the results of RQ3 show that GPT-3.5 without a persona outperforms GPT-3.5 with a persona).
First, we use the prompt design that a single instruction is broken into smaller steps\footnote{https://learn.microsoft.com/en-us/azure/ai-services/openai/concepts/advanced-prompt-engineering?pivots=programming-language-chat-completions\#break-the-task-down}, as depicted in Figure~\ref{fig:prompt-templates-task-decomposition}.
Second, we use the prompt design that contains more detailed instructions\footnote{https://www.promptingguide.ai/introduction/tips}, as depicted in Figure~\ref{fig:prompt-templates-more-detail}.

Table~\ref{tab:result-multi-prompt} shows the results of EM and CodeBLEU that GPT-3.5 achieves across different prompt designs.
The table shows that for zero-shot learning, GPT-3.5 that is prompted by the prompt with a simple instruction achieves 16.44\% - 45.45\% higher EM than GPT-3.5 that is prompted by the prompt with an instruction being broken down into smaller steps.
In addition, GPT-3.5 that is prompted by the prompt with a simple instruction achieves 37.12\% - 880.00\% higher EM than GPT-3.5 that is prompted by the prompt with a detailed instruction.

The table also shows that for few-shot learning, GPT-3.5 that is prompted by the prompt with a simple instruction achieves 5.15\% - 290.00\% higher EM than GPT-3.5 that is prompted by the prompt with an instruction being broken down into smaller steps.
Furthermore, GPT-3.5 that is prompted by the prompt with a simple instruction achieves 5.61\% - 515.79\% higher EM than GPT-3.5 that is prompted by the prompt with detailed instruction.

The results indicate that GPT-3.5 that is prompted by the prompts with a simple instruction achieves the highest EM when compared to other prompt designs.
Thus, the results imply that the prompt with a simple instruction is the most suitable for GPT-3.5 for code review automation.
}

\revised{R1-2}
{
\subsection{Cost and Benefits of Using GPT-3.5 for Code Review Automation}

Cost is one of the factors that determine the choices of AI services for practitioners. 
In the case of GPT-3.5 provided by OpenAI, the cost of using GPT-3.5 varies depending on usage\footnote{https://openai.com/pricing}.
In particular, the cost of using zero-shot learning and few-shot learning with GPT-3.5 is approximately 0.002 USD per query (for 1k input tokens and 1k output tokens) and 0.0035 USD per query (for 4k input tokens and 1k output tokens), respectively. 
On the other hand, the cost for fine-tuning one GPT-3.5 is approximately 40 USD (we use approximately 8k examples from a training set), and the cost for using fine-tuned GPT-3.5 is approximately 0.009 USD per query (for 1k input tokens and 1k output tokens).

Assume that a software developer uses GPT-3.5 to help him/her review code submitted for review 1,000 times per day on average and he/she works for 25 days per month, the total GPT-3.5 usage is 1,000 $\times$ 25 $\times$ 12 = 300,000 times per year.
Such GPT-3.5 usage accounts for \$600 per year (300,000 $\times$ 0.002) when using zero-shot learning with GPT-3.5, \$1,050 per year (300,000 $\times$ 0.0035) when using few-shot learning with GPT-3.5, and \$2,740 per year (40 + (300,000 $\times$ 0.009)) when using fine-tuned GPT-3.5.
However, when compared to the average yearly salary of software engineers around the world\footnote{https://codesubmit.io/blog/software-engineer-salary-by-country/}, the usage cost of GPT-3.5 is approximately 62.23\% - 91.73\% and 97.51\% - 99.46\% less than the lowest average salary  (i.e., \$7,255 in Nigeria) and the highest average salary (i.e., \$110,140 in the United States), respectively.
Nevertheless, the results of RQ1 show that fine-tuned GPT-3.5 achieves the highest EM.
In addition, the results of RQ3 show that the use of few-shot learning without persona on GPT-3.5 helps GPT-3.5 achieve the highest EM and CodeBLEU.
Thus, we recommend that GPT-3.5 for code review automation should be fine-tuned.
Otherwise, leveraging GPT-3.5 by using few-shot learning without persona can be considered as an alternative.
}

\section{Threats to Validity} \label{sec:threat}

We describe the threats to the validity of our study below.

\subsection{Threats to Construct Validity} 
Threats to construct validity relate to the example selection techniques that we use to select examples for few-shot learning, and the design choices of the persona.
We explain each threat below.

In this study, we only use the BM25~\cite{robertson2009probabilistic} technique to select three demonstration examples for few-shot learning.
However, using more demonstration examples or different techniques to select demonstration examples may lead to results that differ from the reported results.
Thus, more demonstration examples and other techniques for selecting demonstration examples can be explored in future work.

Since the code review automation task is related to revising the patches written by software developers, we use the persona (i.e., \textit{an expert software developer}) to ensure that the revised code generated by GPT-3.5 looks like the source code written by a software developer. 
However, there are other similar personas (e.g., a senior software engineer, or a front-end software developer)
that we do not explore.
Thus, future work can explore other design choices of prompt and persona to find the optimal prompt and persona for code review automation tasks.

\subsection{Threats to Internal Validity}
Threats to internal validity relate to the randomness of GPT-3.5 and Magicoder, and the hyper-parameter settings that we use to fine-tune GPT-3.5 and Magicoder.
The results that we obtain from GPT-3.5 and Magicoder may vary due to the randomness of GPT-3.5 and Magicoder.
However, doing the same experiments multiple rounds can be expensive due to large testing datasets.
Finally, we do not explore all possible combinations of hyper-parameter settings (e.g., the number of epoch or learning rate) when fine-tuning GPT-3.5 and Magicoder.
We do not do so since the search space of hyper-parameter settings is large, which can be expensive.
Nonetheless, the main goal of this study is not to find the best hyper-parameter settings for code review automation, but to investigate the performance of GPT-3.5 and Magicoder on code review automation tasks when using model fine-tuning.


\subsection{Threats to External Validity}
Threats to external validity relate to the generalizability of our findings in other software projects.
In this study, we conduct the experiment with the dataset obtained from recent work~\cite{pornprasit2023d, li2022automating, tufano2022using}.
However, the results of our experiment may not be generalized to other software projects.
Thus, other software projects can be explored in future work. 

\revised{R1-8}{
Another threat relates to the updates to GPT-3.5 made by OpenAI in future.
Due to the updates, reproduced experiment results may differ from those reported in this paper.
}

%

\section{Conclusion} \label{sec:conclusion}

In this work, we investigate the performance of LLMs (i.e., GPT-3.5 and Magicoder) for code review automation when using model fine-tuning and inference techniques (i.e., zero-shot learning, few-shot learning, and persona).
We also compare the performance of the LLMs with the existing code review automation approaches~\cite{pornprasit2023d, li2022automating, tufano2022using}.
Our results show that 
(1) fine-tuned GPT-3.5 performs best for code review automation and 
(2) the best prompting strategy when using GPT-3.5 without fine-tuning is few-shot learning without a persona.
Based on the results, we recommend that (1) LLMs for code review automation should be fine-tuned to achieve the highest performance;
and (2) when data is not sufficient for model fine-tuning, few-shot learning without a persona should be used for LLMs for code review automation.




\bibliographystyle{unsrt}
\bibliography{IEEEabrv,ref.bib}

\end{document}